\documentclass[longauth]{aa} 
\usepackage{graphicx}
\pdfoutput=1
\usepackage{txfonts}
\usepackage{natbib}
\bibpunct{(}{)}{;}{a}{}{,}
\begin{document}
   \title{The VIMOS VLT Deep Survey:\thanks{Based on observations collected at the European Southern Observatory New Technology Telescope, La Silla, Chile, program 075.A-0752(A),
 on data obtained with the
 European Southern Observatory Very Large Telescope, Paranal, Chile, program 
 070.A-9007(A), and on observations obtained with MegaPrime/MegaCam, a joint  
project of CFHT and CEA/DAPNIA, at the Canada-France-Hawaii Telescope  
(CFHT) which is operated by the National Research Council (NRC) of  
Canada, the Institut National des Science de l'Univers of the Centre  
National de la Recherche Scientifique (CNRS) of France, and the  
University of Hawaii. This work is based in part on data products  
produced at TERAPIX and the Canadian Astronomy Data Centre as part of  
the Canada-France-Hawaii Telescope Legacy Survey, a collaborative  
project of NRC and CNRS.}}

   \subtitle{The $K$-band follow-up in the $0226-04$ field}
   
   \titlerunning{The VVDS: $K$-band follow-up in the $0226-04$ field}

   \author{S. Temporin \inst{1,2}
\and A. Iovino \inst{2}
\and M. Bolzonella  \inst{4}
\and H.J. McCracken \inst{11,12} 
\and M. Scodeggio \inst{3}
\and B. Garilli \inst{3}
\and D. Bottini \inst{3}
\and V. Le Brun \inst{10}
\and O. Le F\`evre \inst{10}
\and D. Maccagni \inst{3}
\and J.P. Picat \inst{8}
\and R. Scaramella \inst{5,14}
\and L. Tresse \inst{10}
\and G. Vettolani \inst{5}
\and A. Zanichelli \inst{5}
\and C. Adami \inst{10}
\and S. Arnouts \inst{10}
\and S. Bardelli  \inst{4}
\and A. Cappi    \inst{4}
\and S. Charlot \inst{9,11}
\and P. Ciliegi    \inst{4}  
\and T. Contini \inst{8}
\and O. Cucciati \inst{2,15}
\and S. Foucaud \inst{22}
\and P. Franzetti \inst{3}
\and I. Gavignaud \inst{13}
\and L. Guzzo \inst{2}
\and O. Ilbert \inst{21}
\and B. Marano     \inst{7}  
\and C. Marinoni \inst{19}
\and A. Mazure \inst{10}
\and B. Meneux \inst{2,3}
\and R. Merighi   \inst{4} 
\and S. Paltani \inst{16,17}
\and R. Pell\`o \inst{8}
\and A. Pollo \inst{10,19}
\and L. Pozzetti    \inst{4} 
\and M. Radovich \inst{6}
\and D. Vergani \inst{3}
\and G. Zamorani \inst{4} 
\and E. Zucca    \inst{4}
\and M. Bondi \inst{5}
\and A. Bongiorno \inst{7}
\and J. Brinchmann \inst{20}
\and S. de la Torre \inst{10}
\and F. Lamareille \inst{8}
\and Y. Mellier \inst{11,12}
\and C.J. Walcher \inst{10}}

   \offprints{S. Temporin}

   \institute{Laboratoire AIM, CEA/DSM - CNRS - Universit\'e\ Paris Diderot,
DAPNIA/SAp, 91191 Gif sur Yvette, France\\
              \email{sonia.temporin@cea.fr}
\and
INAF-Osservatorio Astronomico di Brera - Via Brera 28, I-20121, Milano, Italy	      
\and
IASF-INAF - via Bassini 15, I-20133, Milano, Italy
\and
INAF-Osservatorio Astronomico di Bologna - Via Ranzani,1, I-40127, Bologna, Italy
\and
IRA-INAF - Via Gobetti,101, I-40129, Bologna, Italy
\and
INAF-Osservatorio Astronomico di Capodimonte - Via Moiariello 16, I-80131, Napoli,
Italy
\and
Universit\`a di Bologna, Dipartimento di Astronomia - Via Ranzani,1,
I-40127, Bologna, Italy
\and
 Laboratoire d'Astrophysique de Toulouse/Tabres (UMR5572), CNRS, 
Universit\'e Paul Sabatier - Toulouse III, Observatoire Midi-Pyr\'en\'ees, 14 
av. E. Belin, F-31400 Toulouse (France)
\and
Max Planck Institut fur Astrophysik, 85741, Garching, Germany
\and
Laboratoire d'Astrophysique de Marseille, UMR 6110 CNRS-Universit\'e de
Provence,  BP8, 13376 Marseille Cedex 12, France
\and
Institut d'Astrophysique de Paris, UMR 7095, 98 bis Bvd Arago, 75014
Paris, France
\and
Observatoire de Paris, LERMA, 61 Avenue de l'Observatoire, 75014 Paris, 
France
\and
Astrophysical Institute Potsdam, An der Sternwarte 16, D-14482
Potsdam, Germany
\and
INAF-Osservatorio Astronomico di Roma - Via di Frascati 33,
I-00040, Monte Porzio Catone,
Italy
\and
Universit\'a di Milano-Bicocca, Dipartimento di Fisica - 
Piazza delle Scienze, 3, I-20126 Milano, Italy
\and
Integral Science Data Centre, ch. d'\'Ecogia 16, CH-1290 Versoix
\and
Geneva Observatory, ch. des Maillettes 51, CH-1290 Sauverny, Switzerland
\and
Astronomical Observatory of the Jagiellonian University, ul Orla 171, 
30-244 Krak{\'o}w, Poland
\and
Centre de Physique Th\'eorique, UMR 6207 CNRS-Universit\'e de Provence, 
F-13288 Marseille France
\and
Centro de Astrofísica da Universidade do Porto, Rua das Estrelas,
4150-762 Porto, Portugal 
\and
Institute for Astronomy, 2680 Woodlawn Dr., University of Hawaii,
Honolulu, Hawaii, 96822
\and	
School of Physics \& Astronomy, University of Nottingham, University Park, Nottingham, NG72RD, UK}

   \date{Received ?? ??, 2007; accepted ? ??, ????}

 
  \abstract
   {}
   {We present a new $K_s$-band survey that represents a significant
   extension to the previous wide-field $K_s$-band imaging survey
   within the $0226-04$ field of the VIMOS-VLT deep survey (VVDS).
   The new data add $\sim$ 458 arcmin$^2$ to the previous imaging
   program, thus allowing us to cover a total contiguous area of
   $\sim$ 600 arcmin$^2$ within this field.}
   {Sources are identified both directly on the final $K$-band mosaic
   image and on the corresponding, deep
   $\chi^2-g^{\prime}r^{\prime}i^{\prime}$ image from the CFHT Legacy
   Survey in order to reduce contamination while ensuring us the
   compilation of a truly K-selected catalogue down to the
   completeness limit of the $K_s$-band. The newly determined
   $K_s$-band magnitudes are used in combination with the ancillary
   multiwavelength data for the determination of accurate photometric
   redshifts.}
   {The final catalogue totals $\sim$ 52\,000 sources, out of which
   $\sim$ 4400 have a spectroscopic redshift from the VVDS first epoch
   survey. The catalogue is 90\% complete down to $K_{\rm{Vega}}$ =
   20.5 mag. We present $K_s$-band galaxy counts and angular correlation 
   function measurements down to such magnitude limit. 
   Our results are in good agreement with previously published work. We
   show that the use of $K$ magnitudes in the determination of
   photometric redshifts significantly lowers the incidence of
   catastrophic errors. The data presented in this paper are publicly
   available through the CENCOS database.}
   {}


   \maketitle
%

\section{Introduction}

 Near-infrared galaxy surveys are widely recognized to give a number
 of advantages with respect to optical surveys as tools to study the
 process of mass assembly at high redshifts.  The observed K-band
 gives a measure of the rest-frame optical fluxes for intermediate
 redshift galaxies up to $z\, \sim$ 2, therefore it can be more easily
 related to the galaxy mass in stars. Furthermore, the K-band
 selection leads to the inclusion of extremely red objects that would
 be otherwise missed by a selection in the optical regime.  Additional
 advantages with respect to optical surveys are given by the smaller
 effects of extinction on K-band observations and by the smaller
 required $k$-correction, with little dependence on the galaxy types.
 Near-infrared data are also particularly useful for a more accurate
 determination of photometric redshifts, a key issue especially in the
 redshift range 1 $<\, z \, <$ 2, where the measurement of
 spectroscopic redshifts can be challenging.
  
 In fact, in recent years several efforts have been devoted to the
 compilation of K-selected samples of galaxies, including
 spectroscopic surveys of K-selected sources such as the K20 survey
 \citep{cimatti02}.  However, surveys reaching very faint K-band
 magnitudes tend to be limited to rather small sky areas (and thus are
 affected by cosmic variance, \citet[e.g. FIRES][down to a depth $K \, \sim$ 24 over a few square arcminutes]{labbe03}), while surveys on
 large areas are often only moderately deep \citep[e.g.][down to $K \, \sim$ 19 over several hundred square arcminutes]{daddi00,drory01,kong06}. 
  A recent overview of the depth
 and area of published near-IR imaging surveys can be found in
 figure~15 of \citet{fs06}.
  
 Only very recently surveys that cover the intermediate regime of area
 and depth have started to appear. Some examples are the UKIDSS Ultra
 Deep Survey and Deep ExtraGalactic Survey \citep[DR1; various levels of depth and coverage over several square degrees,][]{warren07}, the
 MUSYC survey \citep[5$\sigma\, K_{\rm{Vega}}\, \sim\, 21$ over 4 fields 10\arcmin $\times$ 10\arcmin each,][]{quadri07}, and the
 Palomar Observatory Wide-Field Infrared Survey \citep[1.53 deg$^2$ over 4 fields, down to $K_{\rm{Vega}}\, \sim\, 20.5$,][]{conselice07}.
 
 Here we present a new $K$-band survey that, in the context of the
 VIMOS-VLT deep survey \citep[VVDS][]{olf05}, enlarges significantly
 the area already surveyed by \citet{iovino05}, although to a
 shallower depth, thus yielding a wide $K$-band contiguous field of
 $\sim$ 629 arcmin$^2$ within the $0226-04$ VVDS field (F02).  This
 dataset benefits from the multiwavelength information already
 available for the field F02, namely optical and ultraviolet imagery
 available through the VVDS \citep[$UBVRI$; ][]{hjmc03,rado04} and the
 CFHT Legacy Survey
 ($u^{\ast}g^{\prime}r^{\prime}i^{\prime}z^{\prime}$), $J$-band
 imagery available for a sub-area of $\sim$ 161 arcmin$^2$
 \citep{iovino05}, and, in the radio regime, 2.4 GHz VLA and 610 MHz
 GMRT data \citep{bondi03,bondi07a}.  Spectroscopic observations from
 the first epoch VVDS \citep{olf05} targeted 2823 sources from our
 $K$-band catalogue to $K_{\rm{Vega}}$ $\leq$ 20.5.  Limiting
 magnitudes (intended as 50\% completeness limits) for the available
 multi-wavelength imagery are: $u^{\ast}_{\rm{AB}}\,\sim\, 26.5,
 g^{\prime}_{\rm{AB}}\,\sim\, 26.4, r^{\prime}_{\rm{AB}}\,\sim\, 26.1,
 i^{\prime}_{\rm{AB}}\,\sim\,25.9, z^{\prime}_{\rm{AB}}\,\sim\, 25.0$
 \citep{oi06}; $U_{\rm{AB}}\,\sim\, 25.3$ \citep{rado04},
 $B_{\rm{AB}}\,\sim\, 26.5, V_{\rm{AB}}\,\sim\, 26.2,
 R_{\rm{AB}}\,\sim\, 25.9, I_{\rm{AB}}\,\sim\, 25.0$ \citep{hjmc03};
 $J_{\rm{AB}}\,\sim\, 24.1$ \citep{iovino05}.

 The primary aim of this paper is to describe in detail the
 preparation of our $K$-band catalogue and to quantify its reliability
 and completeness.  While the data reduction described in the
 following concerns only the newly obtained, shallower $K$-band data,
 the analysis of the properties is carried out on the entire $K$-band
 catalogue, which includes the deep part of the survey already
 presented by \citet{iovino05} and can be considered complete down to
 $K_{\rm{Vega}} \, \leq$ 20.5, as it is shown below.  Down to this
 magnitude, the deep part of the survey makes up 26\% of the
 catalogue.
 
 The photometric sample and the spectroscopic sub-sample whose general
 properties are described in this paper are then used in a companion
 paper for the selection and analysis of samples of objects with
 extreme colors \citep{st07}. The $K$-band data described here and in
 \citet{iovino05} have significantly contributed to improve the
 determination of the galaxy stellar mass function from the VVDS
 survey, especially for redshifts $z\, >$ 1.2 (up to $z$ = 2.5), and
 for the low-mass tail of the function at lower redshifts, $z\, <$ 0.4
 \citep{pozzetti07}.  Additionally, this $K$-band catalogue is well
 suited to follow the evolution of the rest-frame $I$-band galaxy
 luminosity function \citep{bolzonella07}.  The $K$-band photometry
 presented in this paper has been made available to the astronomical
 community through the CENCOS database \citep{lebrun07} at the URL
 http://cencosw.oamp.fr/, from where it can be retrieved together with
 the photometry in all other available bands (i.e. VVDS photometry in
 UBVRI(J) and CFHTLS photometry in
 $u^{\ast}g^{\prime}r^{\prime}i^{\prime}z^{\prime}$.). The K-band
 catalogue we present here can be easily obtained by quering the
 database with the appropriate K magnitude limits.


\section{Observations and data reduction}

\begin{figure}
   \centering
   \includegraphics[width=8cm,clip=]{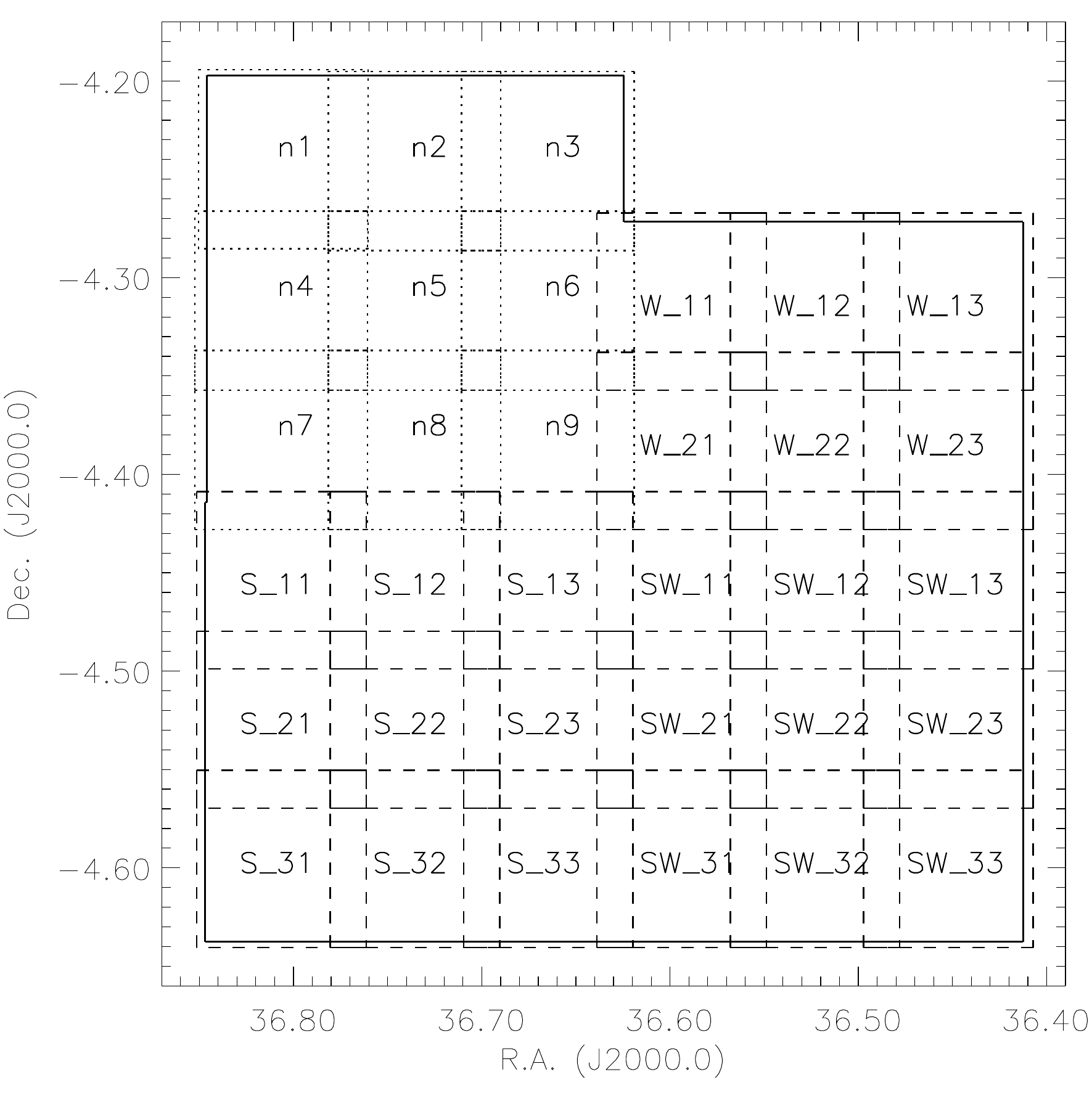}
   \caption{Sky coverage of the $K$-band survey. Dotted lines outline the pointings 
   n1 to n9 of the deep survey \citep{iovino05}. Dashed lines outline the 24 pointings in the
   newly surveyed area, labelled as in Table~\ref{exptimes}. The solid line indicates the 
   final mosaicked area after excision of the borders.}
              \label{pointing-scheme}%
    \end{figure}

New ancillary data to the VVDS 0226-04 field (hereafter F02) have been
obtained in the $K_s$ filter with the SOFI Near Infrared Imaging
camera \citep{moorwood98} at the ESO New Technology Telescope in
September 2005 and February 2006.  These observations cover the region
of sky immediately adjacent to that previously targeted by deep J and
$K_s$ observations \citep{iovino05}, with some overlap on the western
and southern sides.  Hereafter, we refer to the $K_s$-band simply as
$K$-band.  The observations were done in a series of pointings in a
raster configuration, organized in a way to ensure significant overlap
between adjacent pointings as illustrated in
Fig.~\ref{pointing-scheme}, in analogy to the observation scheme
described in \citet{iovino05}.  A total of 24 fields $\sim$
5\arcmin$\times$5\arcmin\ in size, with overlapping borders, have been
observed with a series of jittered 90 s exposures, each obtained with
a detector integration time DIT=10s, and a number of such integrations
NDIT=9. Jittering was performed by randomly offsetting the telescope
within a 30\arcsec\ $\times$ 30\arcsec\ box for a total typical
exposure time of 1 hour per pointing (except for one field which was
exposed for 1.7 hours) with an average seeing of $\sim$ 1\farcs1 and a
pixel scale of 0.288 arcsec pixel$^{-1}$ 
(see Table~\ref{exptimes}
for a list of the pointings and the seeing conditions during the observations).    
The airmass of the data ranges from 1.11 to 1.38, except for part of one pointing that 
was observed in February 2006 at an airmass $\sim$ 1.6.

After excision of the low signal-to-noise borders and of the regions
around bright stars (or very nearby extended galaxies) the new
$K$-band images resulted in a newly covered area of 458.2 arcmin$^2$,
and, when combined with the previous, deeper observations presented in
\citet{iovino05}, in a total $K$-band area of 623 arcmin$^2$ within
F02.  Hereafter, we refer to the combined deep and shallow $K$-band
images as to the K-wide image.

Photometric standard stars from \citet{persson} were observed 5 to 9
times per night. Each standard star was centered within each quadrant
and near the center of the detector array through a pre-defined
sequence of 5 pointings (DIT=1.2s, NDIT=15 each).

%


\subsection{Data reduction}

The data reduction procedure was largely similar to that detailed in \citet{iovino05}.
The individual frames were corrected for dark current, flat-fielded, and examined for
quality assessment by use of IRAF\footnote{IRAF is distributed by the National Optical
Astronomy Observatories, which are operated by the Association of
Universities for Research in Astronomy, Inc., under cooperative
agreement with the National Science Foundation.} packages and scripts. 
The quality check involved the evaluation of fluctuations of the sky-background and 
of the magnitude and the FWHM of a reference star along each exposure sequence, as
well as a check for the presence of artifacts that could compromise the quality of
the final coadded image.
Low-quality images, or images whose seeing was significantly worse compared to that 
of the other images in the same sequence were discarded.
The final total exposure times and the number of useful frames per pointing and observing date, 
after the exclusion of lower quality frames, are listed in Table~\ref{exptimes}, where the 
individual pointings are named according to the scheme of Fig.~\ref{pointing-scheme}.
The average seeing as measured on the final coadded images  and the 
photometric correction terms (see below) applied to each field are reported, too.

The IRDR stand-alone software \citep{sabbey01} was used for the
sky-background subtraction, the coaddition of the images that compose
each exposure sequence, and the construction of the associated weight
maps.  When the exposure sequence of a given pointing was splitted
between two different nights, two separate coadded images were built,
one per night of observation. Exposure sequences were splitted during
the coaddition phase also in the case of a sudden change in the sky
conditions during the observations (e.g. changed background-sky level
and/or seeing).

The reduction of the photometric standard stars proceeded in a similar
way, including the quality assessment of the individual frames.  A
photometric zero-point was determined for each night of observation
through the measurement of the aperture magnitudes of the standard
stars observed during the same night and their comparison with
published magnitudes.  Uncertainties in the zero-points are of 0.01 -
0.02 mag, depending on the night of observation, for the run of
September 2005, and 0.03 mag for the two observing nights of February
2006.  As a value for the atmospheric extinction coefficient we
adopted the one available from the European Southern Observatory web
site for the nearest dates to our observations\footnote{See SOFI web
page: www.ls.eso.org/lasilla/sciops/ntt/sofi/index.html}.  Thus, the
values we used are 0.05 for the 2005 observations and 0.03 for the
observations in February 2006.

The coadded images of F02 were reported to flux values above the
atmosphere and calibrated with the zero-points that compete to their
nights of observation.  Finally the images were rescaled to an
arbitrary common zero-point and reported to the AB magnitude system
with zero point 30.0.  Since not all our nights were of excellent
photometric quality, some refinement of the photometric calibration
was necessary.  To ensure a homogeneous calibration across the final
mosaicked image, the measured magnitudes of the stars in the
overlapping regions of adjacent pointings and in multiple coadded
images of the same pointing (obtained either in different nights or
with different sky conditions), were used to test the photometric
calibration and to determine the flux scale factors to be applied to
the individual coadded images, when necessary.  For this purpose, we
took as reference for the photometry the pointing which was observed
under the best conditions during the observing run, as it emerged from
our quality checks.  This was the pointing SW\_21. The zero-point
uncertainty for that night of observation of SW\_21 is of 0.01 mag.
Scaling factors were ignored both in the case of very small
corrections ($<$ 0.01 mag) and in the case of errors in the
determination of the scale factor larger than the correction
itself. No scaling factor was applied to 14 out of 35 coadded frames.
The correction terms applied to the remaining frames are in the range
0.01 - 0.18 mag, with a median value of 0.08 mag; the individual
values with their uncertainties are given in Table~\ref{exptimes}.
Finally, after the application of these scale factors, the stars in
the overlapping regions between the new $K$-band observations and the
previous deep observations were used to check the consistency of the
absolute photometric calibration with that obtained for the deep-K
survey. No systematic shift was detected in the photometric
calibration.

The astrometric calibration of the coadded images was a two-step
process.  A first order astrometric solution was found taking as
reference the astrometric catalogue of the United States Naval
Observatory (USNO)-A2.0 \citep{monet98}, then the solution was refined
by using as reference the position of non-saturated point-like sources
in the astrometrised CFHTLS $i^{\prime}$-band image of the field.
This procedure, already applied to the deep-K \citep{iovino05}, BVR
\citep{hjmc03}, and U \citep{rado04} images, allowed us to reach a
higher relative astrometric accuracy and to match sources at the
sub-pixel level between optical and K bands.  The rms of these
astrometric solutions ranges between 0\farcs056 and 0\farcs123, with a
median value of 0\farcs099.  The astrometrised and photometrically
calibrated images of the individual pointings, weighted by the
relevant weight maps, were combined in a mosaic together with the deep
$K$-band images and, at the same time, regridded to a final pixel
scale of 0.186 arcsec pixel$^{-1}$, in order to match the scale of the
CFHTLS optical images.  During this operation, performed with the
software SWarp\footnote{Software developped by E. Bertin and available
at the URL http://terapix.iap.fr} \citep{bert02}, the relevant flux
scale factors, determined as explained above, were applied to the
individual coadded images to bring the photometry of the final mosaic
to a homogeneous basis.  A similarly constructed mosaic of the weight
maps was obtained for use in the source detection phase. See
\citet{hjmc03} for further details on the stacking procedure with
SWarp.  The median seeing measured on the K-wide image as the FWHM of
non-saturated point-like sources is $\sim$ 1\farcs0.

\begin{table*}
\begin{minipage}[t]{\textwidth}
\caption{Field F02: Observations summary}          
\label{exptimes}      
\centering                        
\renewcommand{\footnoterule}{}  
\begin{tabular}{c c c c c c l l}        
\hline\hline                 
Pointing & RA & Dec & Obs date & coadded frames & t$_{\rm{exp}}$ & seeing & correction term\footnote{Photometric correction terms that have been applied to bring the photometry into agreement with the reference field SW\_21. }\\
 & $^h \, ^m \, ^s$ & \degr\ \arcmin\ \arcsec\ & & & (min.) & (arcsec) & (mag)\\
\hline
S\_11& 02 27 14 & -04 27 03 & 6 Sept. 05   & 40 & 60 & 1.11$\pm$0.14 & 0.032$\pm$ 0.014\\
S\_12& 02 26 57 & -04 27 03 & 5 Sept. 05   & 40 & 60 & 0.67$\pm$0.09 & 0.032$\pm$ 0.015\\
S\_13& 02 26 40 & -04 27 03 & 6 Sept. 05   & 40 & 60 & 0.69$\pm$0.09 & 0.032$\pm$ 0.025\\
S\_21& 02 27 14 & -04 31 18 & 5 Sept. 05   & 40 & 60 & 1.09$\pm$0.14  & ...\\
S\_22& 02 26 57 & -04 31 18 & 5 Sept. 05   & 7 & 10.5 & 1.11$\pm$0.10 & ...\\
S\_22& 02 26 57 & -04 31 18 & 5 Feb.  06   & 30 & 45 & 1.03$\pm$0.10  & 0.050$\pm$0.035\\
S\_22& 02 26 57 & -04 31 18 & 6 Feb.  06   & 31 & 46.5 & 0.96$\pm$0.08  & 0.050$\pm$0.035\\
S\_23& 02 26 40 & -04 31 18 & 6 Sept. 05   & 40 & 60 & 0.79$\pm$0.13  & ...0\\
S\_31& 02 27 14 & -04 35 33 & 4 Sept. 05   & 40 & 60 & 0.99$\pm$0.14 & 0.175$\pm$0.055\\
S\_31& 02 27 14 & -04 35 33 & 7 Sept. 05   & 10 & 15 & 0.88$\pm$0.06  & ...\\
S\_32& 02 26 57 & -04 35 33 & 4 Sept. 05   & 39 & 58.5 & 1.01$\pm$0.12  & 0.014$\pm$0.010\\
S\_33& 02 26 40 & -04 35 33 & 4 Sept. 05   & 40 & 60 & 0.94$\pm$0.07  & ...\\
SW\_11& 02 26 23 & -04 27 03&7 Sept. 05   & 41 & 61.5 & 0.88$\pm$0.09  & ...\\
SW\_12& 02 26 06 & -04 27 03 &9 Sept. 05   & 40 & 60 & 1.09$\pm$0.05 & ... \\
SW\_12& 02 26 06 & -04 27 03 & 12 Sept. 05 & 10 & 15 & 0.91$\pm$0.15  & ...\\
SW\_13& 02 25 49 & -04 27 03 & 11 Sept. 05 & 20 & 30 & 1.09$\pm$0.01  & ...\\
SW\_13& 02 25 49 & -04 27 03 & 12 Sept. 05 & 30 & 45 & 1.14$\pm$0.18 & ...\\
SW\_21& 02 26 23 & -04 31 18 & 7 Sept. 05  & 40 & 60 & 1.19$\pm$0.09  & ...\\
SW\_22& 02 26 06 & -04 31 18 & 8 Sept. 05  & 40 & 60 & 1.05$\pm$0.15  & 0.078$\pm$0.021\\
SW\_23& 02 25 49 & -04 31 18 & 12 Sept. 05 & 21 & 31.5 & 1.15$\pm$0.16  & -0.142$\pm$0.027\\
SW\_23& 02 25 49 & -04 31 18 & 12 Sept. 05 & 19 & 28.5 & 1.06$\pm$0.02  & 0.078$\pm$0.017\\
SW\_31& 02 26 23 & -04 35 33 & 7 Sept. 05  & 39 & 58.5 & 0.81$\pm$0.08  & ...\\
SW\_32& 02 26 06 & -04 35 33 & 8 Sept. 05  & 39 & 58.5 & 0.85$\pm$0.05  & 0.102$\pm$0.021\\
SW\_33& 02 25 49 & -04 35 33 & 12 Sept. 05 & 34 & 51 & 0.94$\pm$0.10  & ...\\
SW\_33& 02 25 49 & -04 35 33 & 9 Sept. 05  & 10 & 15 & 1.05$\pm$0.05  & ...\\
W\_11& 02 26 23 & -04 18 33 & 6 Sept. 05   & 20 & 30 & 0.85$\pm$0.14  & -0.095$\pm$0.020\\
W\_11& 02 26 23 & -04 18 33 & 8 Sept. 05   & 20 & 30 & 1.52$\pm$0.15  & -0.116$\pm$0.035\\
W\_12& 02 26 06 & -04 18 33 & 9 Sept. 05   & 40 & 60 & 1.40$\pm$0.10 & -0.116$\pm$0.035\\
W\_12& 02 26 06 & -04 18 33 & 12 Sept. 05  & 10 & 15 & 0.92$\pm$0.10  & -0.134$\pm$0.040\\
W\_13& 02 25 49 & -04 18 33 & 11 Sept. 05  & 40 & 60 & 1.21$\pm$0.10  & -0.116$\pm$0.035\\
W\_13& 02 25 49 & -04 18 33 & 12 Sept. 05  & 10 & 15 & 1.10$\pm$0.15  & -0.179$\pm$0.048\\
W\_21& 02 26 23 & -04 22 48 & 8 Sept. 05   & 40 & 60 & 1.16$\pm$0.12  & -0.047$\pm$0.020\\
W\_22& 02 26 06 & -04 22 48 & 9 Sept. 05   & 40 & 60 & 1.19$\pm$0.07  & -0.078$\pm$0.022\\
W\_22& 02 26 06 & -04 22 48 & 12 Sept. 05  & 10 & 15 & 1.18$\pm$0.07  & -0.190$\pm$0.040\\
W\_23& 02 25 49 & -04 22 48 & 11 Sept. 05  & 40 & 60 & 1.08$\pm$0.12  & -0.073$\pm$0.049\\
\hline                                  
\end{tabular}
\end{minipage}
\end{table*}

\subsubsection{Quality of astrometric calibration}

\begin{figure*}
   \centering
   \vbox{
   \includegraphics[width=7.5cm,clip=]{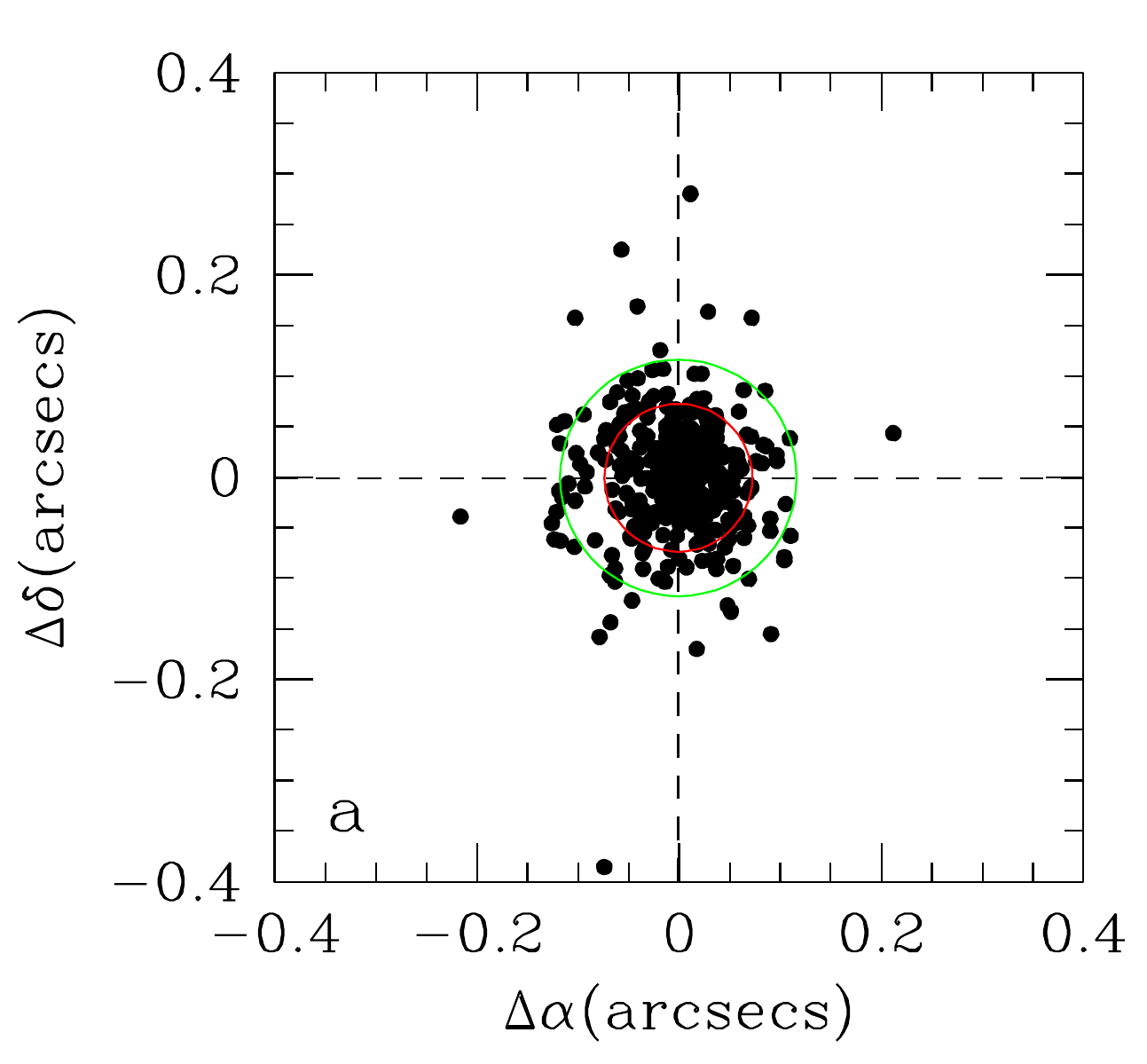}
   \includegraphics[width=8cm,clip=]{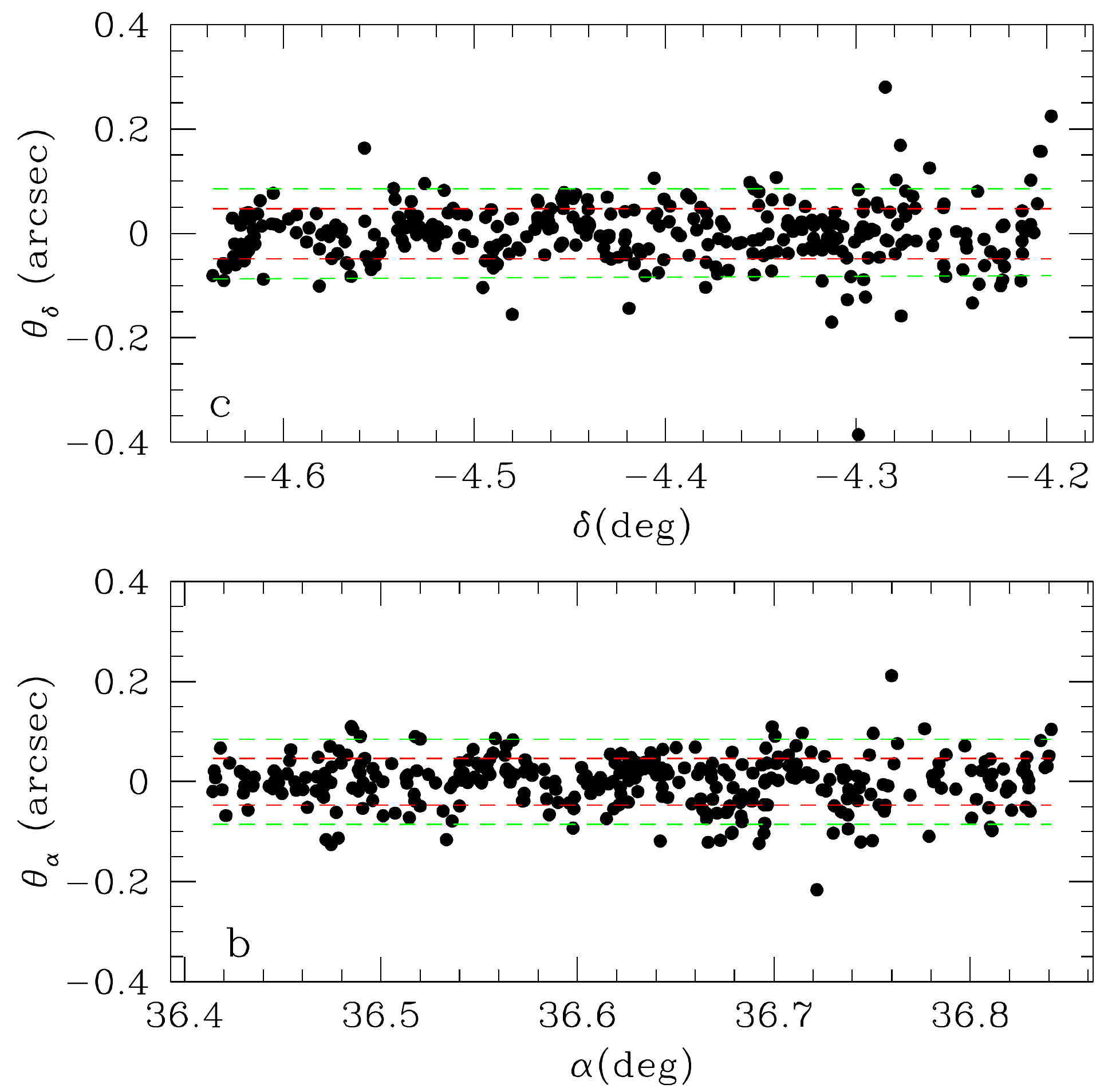}}
   \caption{Radial (a), right-ascension (b), and declination (c) residuals between 
   the $K$-band and 
   $i^{\prime}$-band positions for unsaturated, point-like sources. The inner (red) and outer (green) dashed
   circles/lines enclose 68\% and 90\% of all objects, respectively. The centroid of the
   distribution of radial residuals is indicated by crossed dashed lines.}
              \label{astrometry}%
    \end{figure*}

A catalogue of CFHTLS $i^{\prime}$-band point-like sources was used to
verify the goodness of the astrometric calibration of the final
$K$-band mosaic.  We run SExtractor \citep{ba96} on the image as a
detection algorithm.  A representation of the radial offsets of
K-detected stars with respect to their catalogued positions is shown
in Fig.~\ref{astrometry}.  No systematic offset has been detected
either in right ascension or in declination.  The measured positions
agree with the catalogued ones within 0\farcs07\ (0\farcs12)\ for 68\%
(90\%) of the sources.  A separate examination of the residuals
between $K$-band and $i^{\prime}$-band positions as a function of
right ascension and declination (Fig.~\ref{astrometry}) does not show
any dependence on the location in the frame. The quality of the
astrometric calibration is uniform throughout the final mosaicked
image.  The $K$-band positions agree with the $i^{\prime}$-band ones
within 0\farcs047 (0\farcs086) for 68\% (90\%) of the point-like
sources in both right ascension and declination.
 
\subsubsection{Quality of photometry}

The quality checks on the image sequences of the target and standard
star fields described above, and the comparison of the magnitudes
between the deep and the shallow parts of our $K$-band survey already
gave an indication of the reliability of our photometry.

\begin{figure}
   \centering
   \includegraphics[width=8cm,clip=]{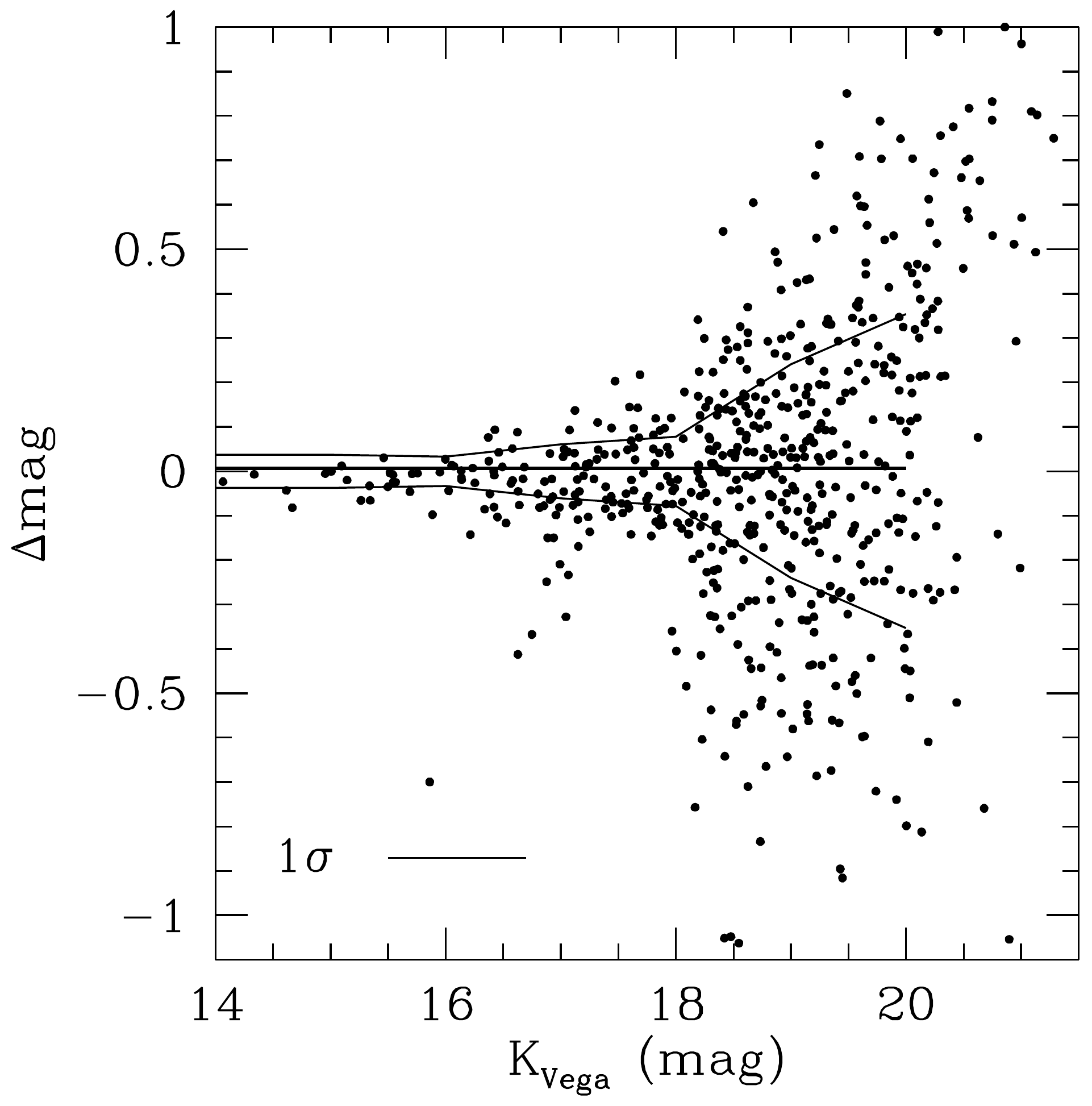}
   \caption{Photometric error as a function of the $K$-band magnitude obtained
   as comparison of the magnitudes from two 30 min exposures of the same field.
   The thick solid line indicates the mean magnitude difference between the two
   images and the thin solid lines indicate the 1$\sigma$ error as a function of
   the $K_{\rm{Vega}}$ magnitude. The typical 1$\sigma$ error in the magnitudes of 
   the final stacked 1hr exposure will be a half of the error reported in this 
   plot.}
              \label{photom-err}%
    \end{figure}

The photometric errors as a function of the K magnitude were estimated
by use of repeated exposures on the same region of the sky.  The
source detection was performed with SExtractor and the MAG\_AUTO
parameter from the output file was adopted as a measure of the total
magnitudes.  An example is shown in Fig.~\ref{photom-err}, where two
30 minutes exposures have been used for the comparison.  Since both
images contribute equally and independently to the measured error
$\sigma$ shown in Fig.~\ref{photom-err}, the uncertainty for an
individual 30 minutes exposure is given by $\sigma/\sqrt 2$.  Then,
coadding the two images to obtain the final stacked image (we recall
that the typical exposure time for the stacked image is 1 hour)
reduces the individual errors by a further factor $1/\sqrt 2$. Hence,
the final error will be a half of the value measured with the above
comparison.  Therefore, for the final stacked image of a pointing the
estimated 1$\sigma$ errors are 0.04, 0.12, and 0.17 mag at
$K_{\rm{Vega}}$ = 18,19, and 20 mag, respectively.

\begin{figure}
   \centering
   \includegraphics[width=8cm,clip=]{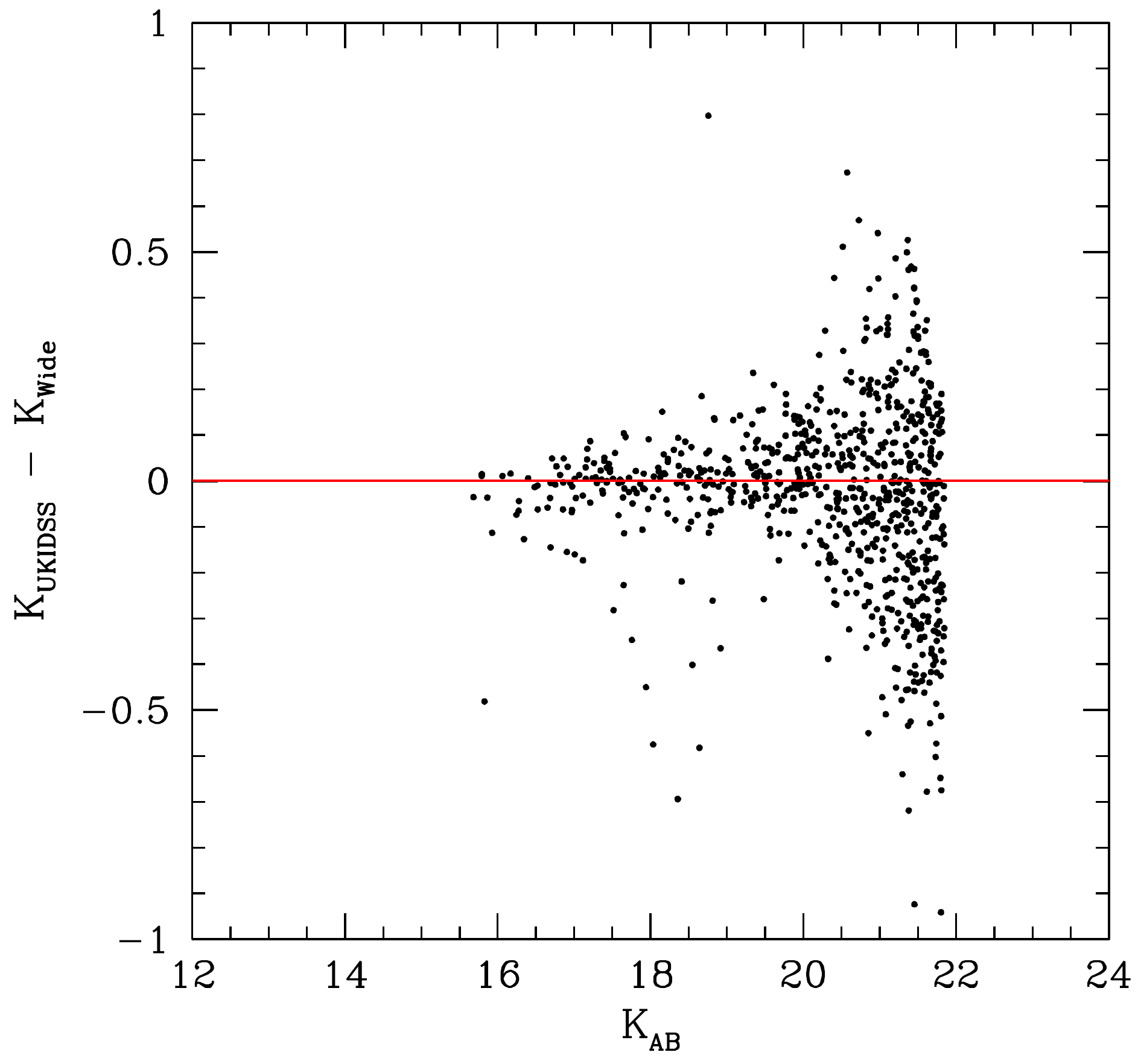}
   \caption{Comparison of our photometry with the $K$-band photometry from the early data release 
   of the UKIDSS survey \citep{dye06}.  Only point-like sources with a 10$\sigma$ detection were
   considered. The solid line indicates the mean value of the magnitude difference.}
              \label{photom-ukidss}%
    \end{figure}

As an independent check of the goodness of our photometry, we compared
our K-band magnitudes with those from the UKIDSS Early Data Release
\citep{dye06} in F02 (Deep Extragalactis Survey, DXS, see figure 14 of
\citet{dye06} for an indication of coverage and depth) and found them
to be in reasonable agreement, although with some spread.  No
significant systematic offset is present.  In Fig.~\ref{photom-ukidss}
we show the comparison for all cross-identified bona-fide point-like
sources that were selected for having a SExtractor FLUX\_RADIUS
parameter $r_{1/2}\, >\, 2.7$ in the $i^{\prime}$-band
\citep[see][]{oi06} and have $K_{\rm{Vega}} \, < 20$ in our catalogue,
roughly corresponding to a $> \, 10\sigma$ detection on our $K$-band
image.

\section{The $K$-band catalogue}

\subsection{Building the catalogue}\label{build-cat}

The construction of the catalogue went through a series of steps aimed
at obtaining a catalogue as complete as possible in the $K$-band --
suited to the extraction of $K$-selected galaxy samples -- while
minimising the incidence of spurious detections without affecting our
completeness.  This goal was achieved by matching the results of two
detection procedures, one performed directly on the $K$-band image,
and the other based on the use of a $\chi^2$ image, constructed using
the technique outlined in \citet{sz99}.  The latter had the main
purpose of pruning spurious detections from the catalogue, especially
in the $K$-faint regime, and to obtain $K$-band magnitudes already
matched with the optical ones, which were obtained starting from the
same $\chi^2$ detection image \citep[see also][for the use of $\chi^2$ images in the preparation of VVDS photometric catalogues in other wavelengths]{hjmc03,rado04}.

The steps we followed are detailed below.  The detection and
measurement of the sources on the final mosaicked image with the
associated weight map were obtained with SExtractor
by applying similar criteria to those described in \citet{iovino05}.
As a measure of total magnitudes in our catalogue we use the
SExtractor parameter MAG\_AUTO.

\emph{Step 1}. A $\chi^2-g^{\prime}r^{\prime}i^{\prime}$ image built
from CFHTLS data has been used as the detection image for the
measurement of the $K$-band sources with SExtractor in double-image
mode. The use of the $\chi^2$ image has the advantage of reducing the
number of spurious detections, while still picking up most reliable
sources that are present in our $K$-band image. The $\chi^2$ image and
the $K$-wide image have sufficiently similar seeing to assure the
effectiveness of this procedure.  For an object to be included in our
catalogue, it must contain at least three contiguous pixels above a
per-pixel detection threshold of 0.4$\sigma$ in the $\chi^2$ image.
This step produced the version 1 (V1) of our catalogue and offered us
a direct cross-identification of the $K$-band sources with those
listed in the CFHTLS catalogue for the same region of the sky
\citep{oi06}. The total number 

\emph{Step 2}. A second version of the catalogue (V2) was produced by
using SExtractor in single-image mode (i.e. without a separated
detection image) on the $K$-wide image.  This was a key step for the
creation of a $K$-selected catalogue and was necessary to ensure that
we would not miss any real $K$-band source that escaped detection on
the $\chi^2$ image.  In this case, to reduce the number of fake
detections, we chose to include in the catalogue the objects
containing at least 9 contiguous pixels above a per-pixel detection
threshold of 1.1$\sigma$. 

\emph{Step 3}. A positional match of sources between the two
catalogues was obtained by adopting 1\farcs0 as an upper limit to the
separation for a source to be considered the same. Tests with
different upper limits to the source separation showed that our choice
was conservative enough to include all sources present in both
catalogues and sufficiently small to minimize false matches. This
procedure provided us with a list of objects that were detected in the
K-wide image but were not in the $\chi^2$ image. Besides truly red
objects, this list included members of (apparent) pairs that are
resolved in the $K$-band image but appear as an individual source in
the $\chi^2$ image (see Step 6 below), sources that fall in the
vicinity of bright objects and are, therefore, masked by bright halos
in the $\chi^2$ image, and, finally, a number of fake detections that
needed to be expunged.

\emph{Step 4}. In order to refine the object list produced at Step 3,
we evaluated the level of contamination from fake detections as a
function of magnitude.  To this purpose we applied SExtractor to the
inverse of the $K$-wide image, with the same parameters we used for
the production of the catalogue V2.  The contamination level within
0.5 magnitude bins was estimated as the fraction of detections in the
inverse image with respect to the original image.  The resulting
contamination rate is 9\% in the bin centered at $K_{\rm{Vega}}$ =
19.75 and reaches 23\% in the bin centered at $K_{\rm{Vega}}$ = 20
(i.e. in the range 19.75 - 20.25 mag).  We note that this
contamination rate is relevant for the sources that were detected
using only the $K$-band image, while it is not representative of the
contamination from spurious detections of catalogue V1, for which the
effect is minimized by the use of the $\chi^2$ image \citep[as demonstrated in ][]{iovino05}.
 
To limit the amount of contamination, out of all $K$-band detections
for which it was not possible to identify a counterpart in the
catalogue V1 we selected those with a magnitude K$_{\rm{Vega}}$ $\leq$
20.25 and, out of these, we retained only the sources that appeared
reliable upon visual inspection of the $K$-band and optical images.
These remaining 726 sources were added to the catalogue V1. Their
magnitudes in the other available bands were determined by running
SExtractor on the relevant optical images by using the $K$-band as the
detection image.

\emph{Step 5}. A positional match of our $K$-band detections
(catalogue V2) with catalogued VVDS sources in the relevant region of
the sky, allowed us to ensure the inclusion in our catalogue of all
objects with a measured spectroscopic redshift.  At this step we added
to our catalogue further 66 K-detected sources.  These sources did not
have a counterpart in the catalogue V1 because of their vicinity to
bright stars in the $\chi^2$ image. Also, they were missed by our Step
4 because most of them are fainter than the threshold $K_{\rm{Vega}}$
= 20.25 we adopted for the inclusion of additional sources from the
catalogue V2.  In fact they are $\ga$ 3$\sigma$ detections with an
average magnitude $K_{\rm{Vega}}\, \sim$ 20.9, 13 of them having
$K_{\rm{Vega}}\, <$ 20.5.

\emph{Step 6}. A comparison of the magnitudes of the sources in common
between the catalogues V1 and V2 showed a general good agreement, thus
indicating that the additional sources identified in catalogue V2
could be safely added to the main catalogue without need of aperture
corrections. However, a number of sources showed a remarkable
magnitude difference between the two catalogues and required further
investigation.  We adopted as a criterion to select sources with
deviant magnitudes the simultaneous satisfaction of the conditions:
$|K_{\rm{Vega\_V1}}\, - \,K _{\rm{Vega\_V2}}| \, > \, 5\sigma$ (where
$\sigma$ was obtained by adding in quadrature the errors on the
individual magnitudes) and $|K_{\rm{Vega\_V1}}\, -
\,K_{\rm{Vega\_V2}}| \, > \, 0.2$ (to exclude the cases with small
absolute deviations but with very small magnitude errors, that would
be selected by our first condition). The search was limited to objects
brighter than $K_{\rm{Vega}}$ = 20.25, in analogy to Step 4.
 
 About 300 objects were selected this way and visually inspected to
 check for correct cross-identification between the two catalogues and
 for the presence of possible problems. These objects were found to
 include false matches, matches with $K$-band false detections, source
 blends that had been correctly deblended only in the $K$-band image
 but not in the $\chi^2$ image, and vice-versa, source blends that had
 been correctly deblended in the $\chi^2$-image but remained
 unresolved as an individual source in the $K$-band image.  We
 corrected the catalogue for the relevant cases, by rejecting false
 detections and adopting magnitude measurements based on the $K$-band
 image instead of the $\chi^2$-image for blends that were unresolved
 in the latter. Vice-versa, for blends that were unresolved in the
 $K$-band but correctly resolved in the $\chi^2$-image we kept the
 identifications and magnitudes of catalogue V1. In both cases the
 fact that these sources were actually blends of multiple sources and
 not multiple clumps within a single galaxy appeared obvious from the
 visual inspection of the individual optical and near-infrared images.

\emph{Step 7}. For the deep area of the survey, a comparison between
the magnitudes obtained using the new $\chi^2$ image and those
presented in \citet{iovino05}, based on a $\chi^2-BVRIK$ image, did
not reveal any systematic difference and showed that the new
measurements are equivalent (within the errors) to the original ones.
Therefore, for consistency with the previously published paper, in our
final catalogue we adopted the original magnitudes for the deep part
of the survey.

\emph{Step 8}. A correction for Galactic extinction on an
object-by-object basis was applied to the magnitudes in the final
catalogue by using \citet{sch98} dust maps.

\subsection{Completeness}\label{compl}

An indicative estimate of the limiting magnitude of our survey can be
simply obtained based on the background rms of the coadded images and
on the seeing during the observations.  The $n\sigma$ magnitude limits
are given by mag($n \sigma$) = $zp\, - \, 2.5 \log (n \sigma \sqrt
A)$, where $zp$ is the zero point and $A$ is the area of an aperture
whose radius is the average FWHM of unsaturated point-like sources
(see Table~\ref{exptimes}). The values estimated this way for $n$ = 3,
5 are mag(3$\sigma$) $\sim$ 21.4 and mag(5$\sigma$) $\sim$ 20.9 (in
the Vega system).  Indeed, by measuring the signal-to-noise ratio as a
function of K-band magnitude across the final K-wide image in a
3\farcm1-wide running box with a step of 37\arcsec (i.e. 200 pixels)
(and by imposing the relevant positional constraints to keep the box
within the borders of the $K$-wide image as marked with a solid line
in Fig.~\ref{pointing-scheme}) we obtained a magnitude limit for
3$\sigma$ detections that varies in the range $K_{\rm{Vega}}$ = 21.4 -
22.8, the brightest limit being referred to the shallower part of the
image, which is the subject of the present paper.

\begin{figure}
   \centering
   \includegraphics[width=8cm,clip=]{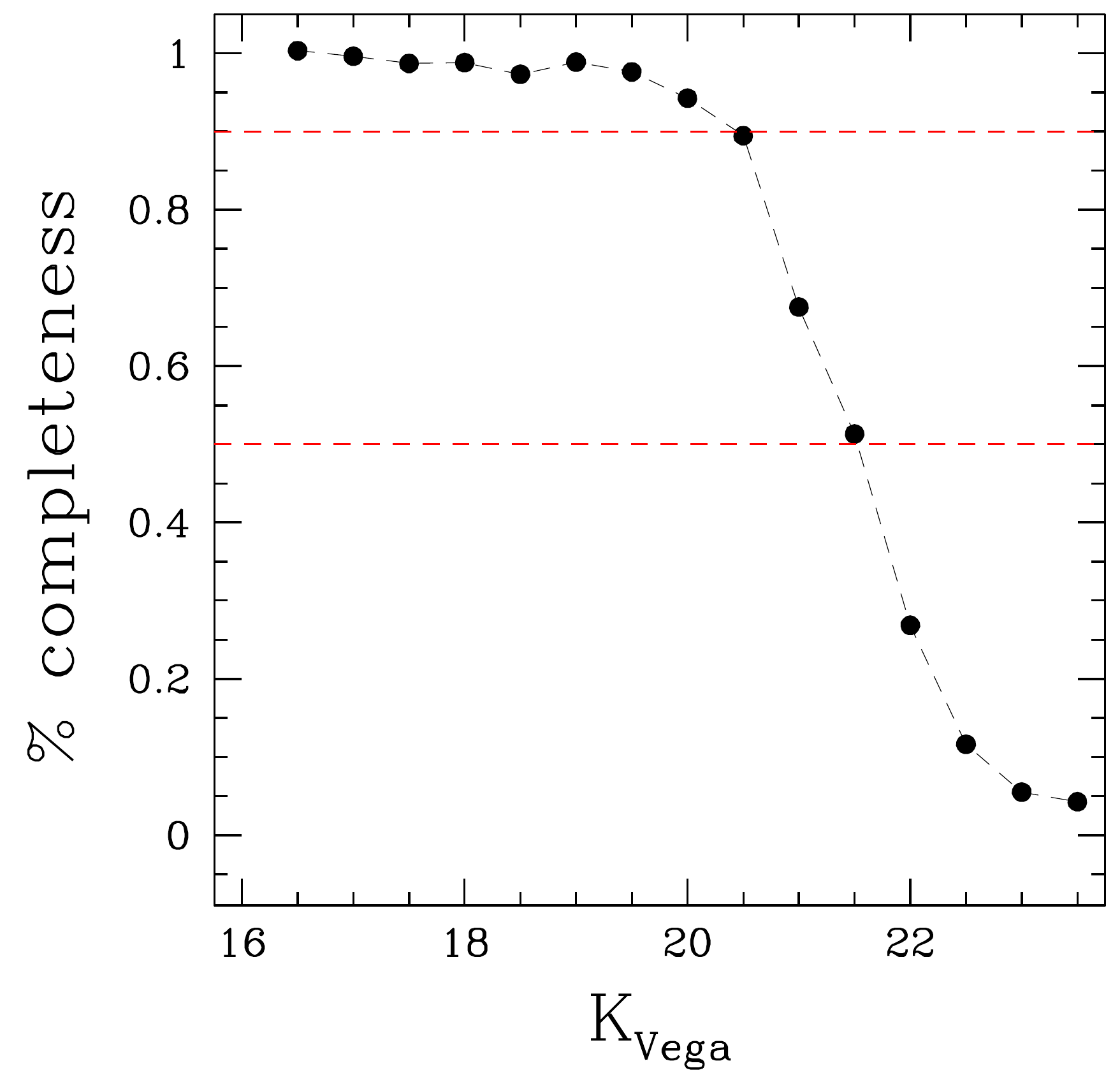}
   \caption{Completeness level of our K-shallow survey as a function of magnitude; 50\% and 90\% completeness levels are marked with dashed lines.}
              \label{completeness}%
    \end{figure}

However, a better characterisation of the photometric properties of
our final image is given by the estimate of the completeness level in
the source detection at various magnitudes. We determined the
completeness level from our capability to recover artificial
point-like sources inserted at random positions in our image.  The
detection of artificial sources was performed with the same SExtractor
parameters adopted for the real sources.  The procedure we followed
does not differ from that adopted by \citet{iovino05} and we refer the
reader to that paper for further details.  A representation of the
completeness level as a function of magnitude for the shallow part of
our $K$-band survey is shown in Fig.~\ref{completeness}. In particular
we reach a nominal completeness level of 90\% (50\%) for point-like
sources with $K_{\rm{Vega}}$ $\leq$ 20.5 (21.5).  The completeness
level for the deep part of the survey \citep[as estimated in][]{iovino05} 
is 90\% (50\%) to $K_{\rm{Vega}}$ $\leq$ 20.75 (22.00).

However, the determination of the actual completeness of our final
catalogue is more complex than described above.  In fact, while the
completeness test was run on the $K$-band image alone, the use of the
$\chi^2$-image played a role in the completeness level that was
finally achieved. On the other hand, having used the
$\chi^2$-image for the detection of sources, there is the concern that
our catalogue is not a purely K-selected one, at least for the shallow
part of the survey that is the object of this paper. As explained in
the previous section, for $K_{\rm{Vega}}\, \leq$ 20.25 all sources
with a significant detection in the $K$-band image have been
included. Furthermore, for $K_{\rm{Vega}}\, \leq$ 20.50 the depth of
the images composing the $\chi^2$-image implies that the last is fully
sufficient to recover sources even with the reddest colors while, at
the same time, reducing the number of spurious detections. Therefore,
our conclusion is that down to $K_{\rm{Vega}}\, =$ 20.5, where we
reach a nominal completeness level of 90\% for point-like sources, we
can safely state that our catalogue is a truly K-selected one.
The achieved completeness level is supported by the raw number counts
(see Sect. 6) that do not fall below a power law at least up 
to $K_{\rm{Vega}}\, =$ 20.5.

Remaining potential problems for the $K_{\rm{Vega}}\, >$ 20.5 regime
are i) possible cases of association of $\chi^2$-detected sources with
correlated noise in the $K$-band image and ii) some possible level of
color incompleteness that could arise from missing extremely red,
$K$-faint sources that could have remained undetected in the $\chi^2$
image. Therefore, for such faint magnitudes regime we cannot state that
our catalog is a purely $K$-selected one, having nevertheless been
designed to be as complete as possible in both the optical and in
K-band. This caveat should be kept in mind when in the next sections
we refer to our catalogue as a $K$-selected one: this attribute is,
strictly speaking, correct only for $K_{\rm{Vega}}\, <$ 20.5.

\subsection{The $K$-wide photometric sample}

Our final $K$-wide catalogue totals 51959 detections. It contains
22846 objects down to our limiting magnitude (i.e. 50\% completeness
limit) $K_{\rm{Vega}} \, \leq 21.5$ and 10615 objects down to our 90\%
completeness limit
\footnote{Hereafter we refer to the 90\% and 50\% limits as to the
``completeness limit'' and the ``limiting magnitude'' of our survey.}
$K_{\rm{Vega}} \,\leq$ 20.5. 

\begin{figure}
\centering
\includegraphics[width=\columnwidth,clip=]{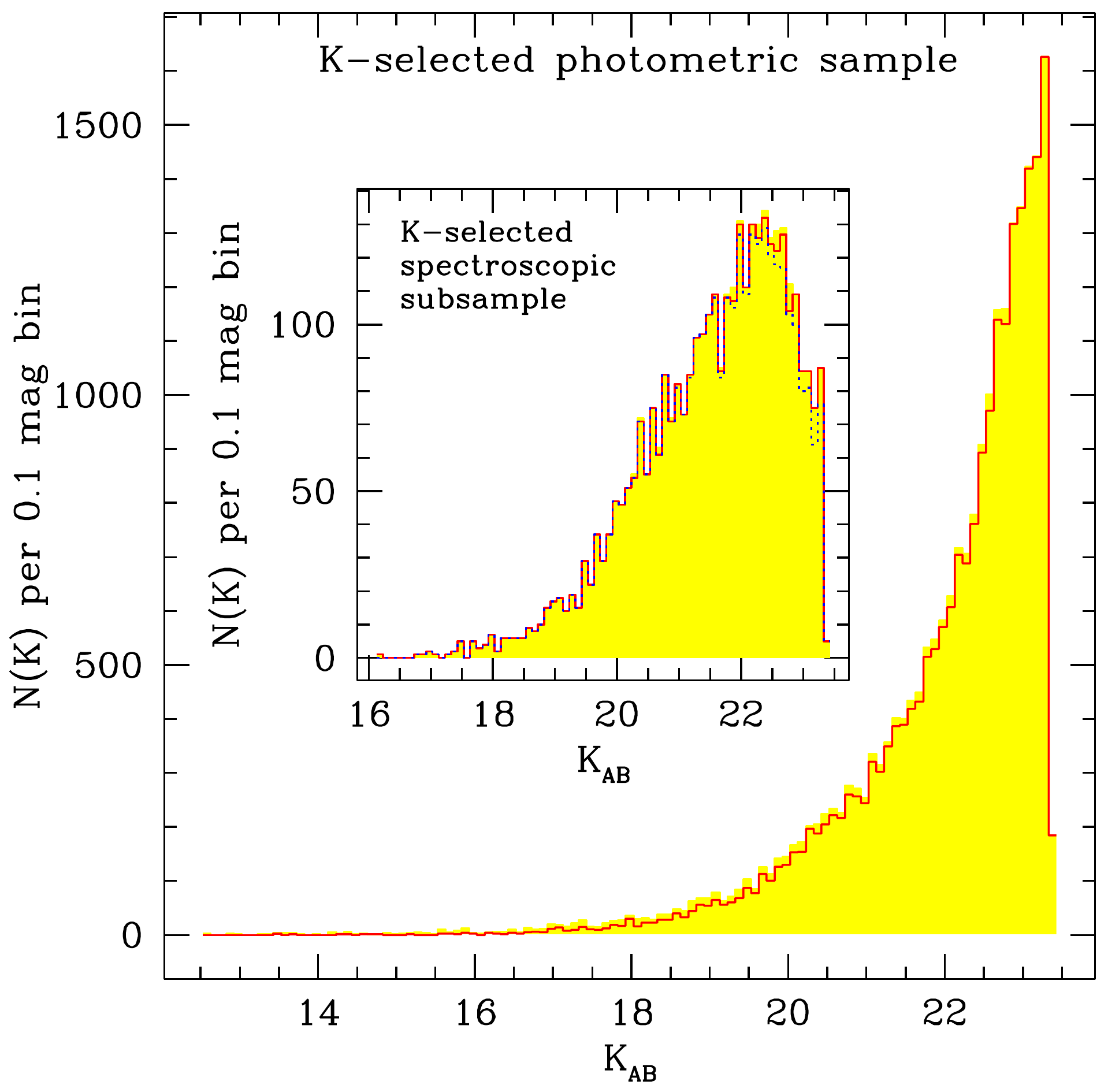}
\caption{Magnitude distribution of our $K$-wide photometric sample down to 
the magnitude limit of the survey $K_{\rm{Vega}} \,\leq$ 21.5 for all objects (yellow
filled histogram)
and for objects classified as galaxies (red histogram; Sect.~5). The inset shows the
 $K$-magnitude
distribution for the whole $K$-selected spectroscopic sample (yellow filled histogram),
 for the objects
classified as galaxies within it (red histogram; Sect.~5), and for the galaxy 
subsample with 
17.5 $\leq\, I_{\rm{AB}}\, \leq$ 24.0 (dotted blue histogram).}
\label{NKmag}
\end{figure}

However, to be sure to be dealing with a truly K-selected
catalogue and to avoid contamination by spurious sources, for future
analysis we restrict our final photometric sample to the 8856 objects
down to $K_{\rm{Vega}} \,\leq$ 20.25.  The number of catalogued
objects for different magnitude limits are reported in
Table~\ref{Nphot_maglim} for ease of comparison with other existing
surveys. At this level, no distinction has been made between stars and
galaxies.  This issue is addressed later on, in Sect.~\ref{sg-sep}.
The magnitude distribution of the sample is shown in Fig.~\ref{NKmag}.
The photometric redshifts for this sample, determined including the
$K$-band information, are presented in Sect.~\ref{photo-z}.

\begin{table}
\begin{minipage}[t]{\columnwidth}
\caption{Field F02: $K$-selected photometric sample}          
\label{Nphot_maglim}      
\centering                        
\renewcommand{\footnoterule}{}  
\begin{tabular}{c c c c c c}        
\hline\hline                 
 & & & & & \\
$K$\footnote{All K magnitudes are expressed in the Vega system. Magnitudes in the AB system 
can be obtained by adding +1.84.}$\leq$ 19.0 & $K\leq$ 19.5& $K\leq$ 19.75& $K\leq$ 20.0& $K\leq$ 20.25 & $K\leq$ 20.5\\
 & & & & & \\
 \hline
 & & & & & \\
3608& 5152& 6180& 6399& 8857 & 10615\\
 & & & & & \\
	\hline
\end{tabular}
\end{minipage}
\end{table}

\subsection{The $K$-selected spectroscopic sample}

Our shallow $K$-band survey and part of the deep one (i.e. all
pointings in Fig.~\ref{pointing-scheme} except for n1, n2, and n3) are
located in the four-pass area of the VVDS, wich implies a
spectroscopic sampling rate up to $\sim$ 40\% down to $I_{\rm AB}\,
\leq$ 24 \citep{olf05}.  Our $K$-wide catalogue (down to the magnitude
limit of the survey, intended as the 50\% completeness limit) contains
a total of 4059 objects covered by spectroscopic observations, out of
which 3815 have a successful measure of redshift (81 of them are
secondary objects with $I_{\rm{AB}}\, > $ 24).  The actual
spectroscopic sampling rate for the $K$-wide catalogue down to its
90\% completeness limit is 26.6\% (27.9\% to $K_{\rm{Vega}}\, \leq$
20.25), while the success rate is $\sim$ 95.5\% down to the same
magnitude limit.  Secure \citep[flags 2, 3, and 4][]{olf05}
spectroscopic redshifts are available for 1792 galaxies and 23 active
galactic nuclei (AGN) down to $K_{\rm{Vega}}\, \leq$ 20.25 (12 of
these objects have $I_{\rm{AB}}\, > $ 24).  The number of galaxies and
stars in the sample within various magnitude limits and for various
levels of spectral quality is summarised in Table~\ref{Nspec_maglim}.
The star-galaxy separation for the spectroscopic sample relies on the
spectral classification.

\begin{table*}
\begin{minipage}[t]{\textwidth}
\caption{Field F02: Number of galaxies/stars in the $K$-selected spectroscopic sample}   
\label{Nspec_maglim}      
\centering
\renewcommand{\footnoterule}{}  
\begin{tabular}{l c c c c c c c c }        
\hline\hline                 
 & & & & & & & & \\
 Flags\footnote{Flags indicate the spectral quality and reliability of redshift measurements
  according to the definitions in \citet{olf05}.}  &
 K\footnote{All K magnitudes are expressed in the Vega system. Magnitudes in the AB system 
can be obtained by adding +1.84. 
In this table the star/galaxy separation is based on a purely spectroscopic classification.}$ \leq$ 19.0 & K $\leq$ 19.5 & K$\leq$ 19.75 & K$\leq$ 19.8 & K$\leq$20.0 & K$\leq$20.25 & K$\leq$20.5 & K$\leq$21.5\\
 & & & & & & & & \\
 \hline
 & & & & & & & & \\
any &	893 / 176&
	1303 / 201&
	1566 / 219&
	1613 / 222&
	1810 / 239&
	2113 / 250&
	2426 / 265&
	3489 / 325\\
1, 21 &	 71 / 7&
	 126 / 11&
	 167 / 12&
	 173 / 12&
	199 / 14&
	251 / 14&
	312 / 15&
	534 / 27\\
2, 22 &	 183/ 11&
	 304 /13 &
	 385 / 14&
	 394 / 15&
	456 / 20&
	551 / 23&
	656 / 27&
	1022 / 52\\
3,4,23,24 &	613 /158 &
	  834/ 177&
	  968/193 &
	  999/ 195&
        1097  / 205&
	1241  / 213&
	1375 / 223&
	1791 / 246\\
9,29 &	 7/ 0&
	 13/ 0&
	 18/ 0&
	 18/ 0&
	25 / 0&
	36 / 0&
	48 / 0&
	105/ \\
11, 211, 12, 212 & 0 / 0 &
	     0 /0 &
	     0 / 0&
	    0 / 0&
           3 / 0&
	   3 / 0&
	   2  / 0&
	   2 / 0 \\
13, 14, 213, 214 & 16 / 0&
	    21 / 0&
	    22 / 0&
	    22 / 0&
           22/ 0 &
	   23 / 0 &
	   25 / 0 &
	   26 / 0 \\
19, 219 &  3 / 0&
	    5 / 0&
	    6 / 0&
	    7 / 0&
	    7 / 0&
	    8 / 0 &
	    8 / 0 &
	    9 / 0 \\
 & & & & & & & & \\
\hline
\end{tabular}
\end{minipage}
\end{table*}

\begin{figure}
\centering
\includegraphics[width=\columnwidth,clip=]{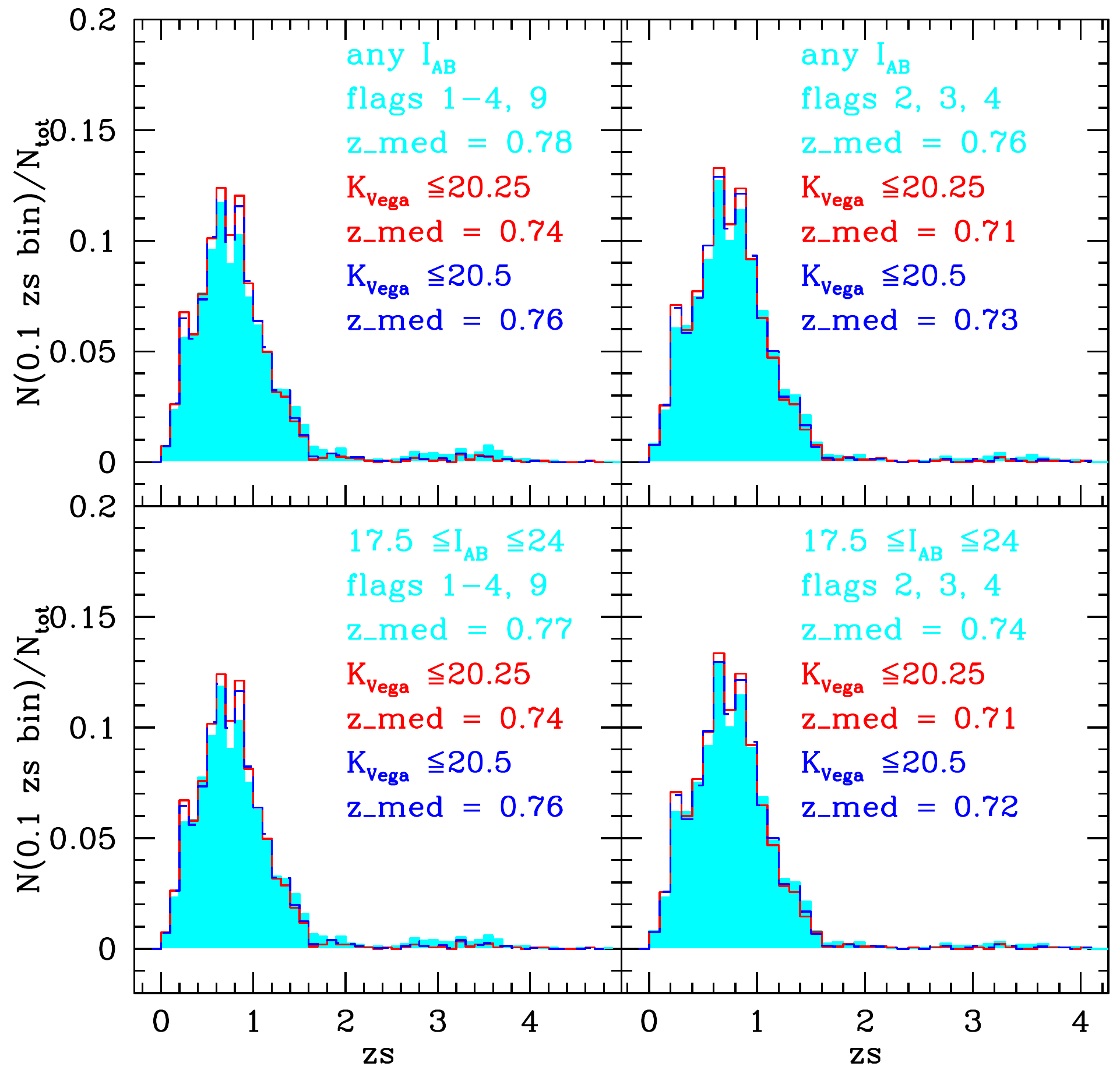}
\caption{Normalized redshift distribution of the $K$-selected spectroscopic sample (including only 
sources spectroscopically classified as galaxies) down to 
$K_{\rm{Vega}}\, \leq$ 20.25 (red empty histogram) and to $K_{\rm{Vega}}\, \leq$ 20.5
(dashed blue line) 
compared to the one of the entire VVDS sample within the same sky area 
(cyan filled histogram).
The left-hand panels include flags 1-4,9, and their equivalents, whereas the right-hand
panels include only secure redshifts with flags 2, 3, 4, 9 and their equivalents.
The top panels show the redshift distribution irrespective of the $I$-magnitude
values (i.e. including secondary objects falling outside the selection criteria
of the spectroscopic survey), whereas the bottom panels show only the objects within the
VVDS magnitude selection limits. The median redshifts $z_{med}$ of the distributions are quoted.}
\label{Nzspec}
\end{figure}

The normalized redshift distribution of our $K$-selected spectroscopic
sample down to the magnitude limits $K_{\rm{Vega}}$ $\leq$ 20.25,
20.50 is shown in Fig.~\ref{Nzspec} in comparison with the redshift
distribution of the whole VVDS sample within the same region of the
sky.  The distribution is presented both for the case of secure
redshifts and for all measured redshifts.  We distinguish the case in
which all available redshifts are considered, irrespective of the
$I$-band magnitudes and the case where only objects within the
magnitude limits of the spectroscopic survey ($17.5\, \leq\,
I_{\rm{AB}}\, \leq\, 24.0 $) are taken into account.  The median
redshifts of the distributions down to $K_{\rm{Vega}}$ $\leq$ 20.5
closely approach those of the whole VVDS samples both for secure and
less secure redshifts (see values quoted in Fig.~\ref{Nzspec}).

The $K$-magnitude distribution of the spectroscopic sample is shown in
Fig.~\ref{NKmag}. The distributions for both the pure $K$-selection
and the combined $K$/$I_{\rm{AB}}$-selection are displayed.

\begin{figure}
   \centering
   \includegraphics[width=9cm,clip=]{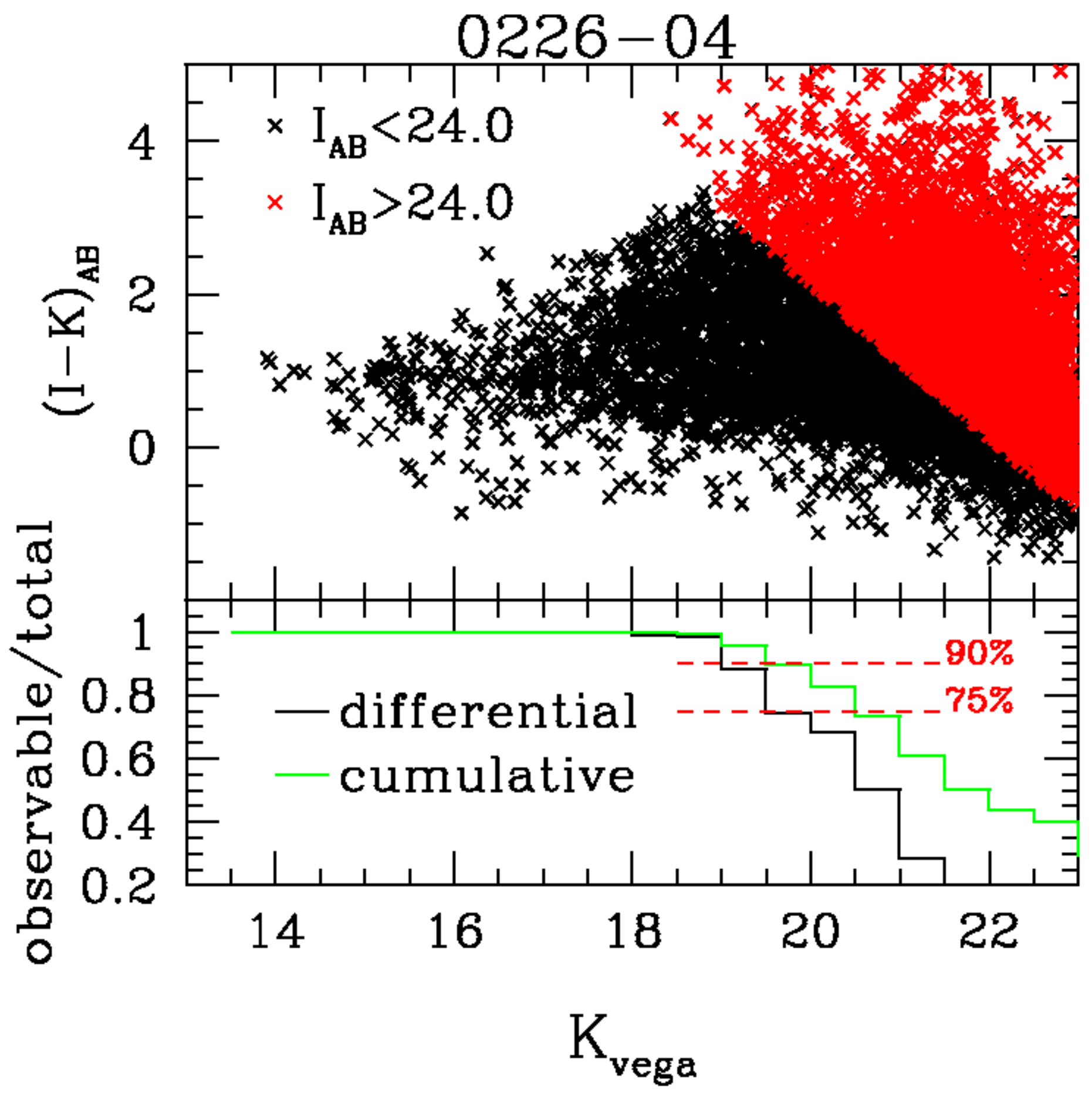}
   \caption{The $I-K$ vs $K$ magnitude diagram for sources with $I_{\rm AB} \, < \,24$ (black)
   and $I_{\rm AB} \, > \,24$ (red) and the differential (black) and cumulative (green) 
   distributions of $K$ magnitudes illustrate the effects of the $I_{\rm AB}$ magnitude cut 
   of the spectroscopic survey on the completeness of the $K$-selected spectroscopic sample.
   A slight colour incompleteness arises in the last two 0.5 mag bins down to $K_{\rm{Vega}}\, 
   \leq \,$ 20.5.}
              \label{col-incomp}
    \end{figure}

A slight incompleteness in colour is present at faint $K$-band
magnitudes to the completeness limit of our survey because of the
$I_{\rm AB} \, \leq \,24$ magnitude limit of the VVDS specroscopic
survey, that tends to disfavour faint red objects. The effect of the
spectroscopic $I_{\rm AB}$ magnitude limit is illustrated in
Fig.~\ref{col-incomp}, where it is shown that for $K_{\rm{Vega}}\,>$
19.0 the spectroscopic survey starts missing some objects with red
$I-K$ colours.  Objects with $I_{\rm AB} \, > \,24$ represent an
increasing fraction of progressively bluer sources for increasingly
faint $K$ magnitudes.  As a result, a $\sim$ 10\% (13\%) colour
incompleteness arises in the 0.5 magnitude bin centered at
$K_{\rm{Vega}}$ = 19.75 (20.25).  However, photometric redshifts can
be used to circumvent this problem.

\section{Photometric redshifts}\label{photo-z}

The photometric redshifts for our $K$-selected sources were obtained with the
code Le\_Phare\footnote{http://www.lam.oamp.fr/arnouts/LE\_PHARE.html} 
(developed by S. Arnouts and O. Ilbert), by following the procedure described
in \citet{oi06} and using MAG\_AUTO magnitudes\footnote{Magnitudes 
were measured in all bands with SExtractor in double-image mode, by using
the $\chi^2$ image as detection image.} measured in the maximum number of 
available photometric bands in the set
$u^{\ast}g^{\prime}r^{\prime}i^{\prime}z^{\prime}BVRIJK$.
$J$-band magnitudes (or upper limits) are available only for a sub-area of 
$\sim$ 160 arcmin$^2$, i.e. for 26\% of the objects in our $K$-band catalogue.
In all our photometric redshift determinations we applied the corrections for
systematic offsets of the zero-points and the optimised galaxy templates 
described in detail in \citet{oi06}.

As it has been shown by \citet{oi06}, a Bayesian approach
\citep{benitez00} with the introduction of an a priori information on
the redshift probability distribution function improves significantly
the determination of photometric redshifts with respect to the
traditional $\chi^2$ method.  In this work, consistently with
\citet{oi06} and supported by the similarity shown in
Fig.~\ref{Nzspec} between $K\, < $ 20.5 and VVDS redshift
distributions, we use directly a prior based on the second one and we
compare the results obtained with/without the application of this
prior to those obtained with/without the use of the $K$-band
photometry in addition to the other bands during the fitting procedure
with Le\_Phare.  This exercise allows us to determine the influence of
the $K$-band on the quality of the photometric redshifts.  In
comparing the photometric redshifts ($zp$) with the spectroscopic ones
($zs$), we express the redshift accuracy in terms of $\sigma_{\Delta
z/(1+zs)}$ using the normalised median absolute deviation defined as
1.48 $\times$ median$(|\Delta z|/(1+zs))$, while for the catastrophic
errors we adopt the definition $|zs\, -\, zp|/(1+zs)\, >\, 0.15$.  In
analogy to \citet{oi06}, we splitted our $K$-selected spectroscopic
sample into an $i^{\prime}$-bright ($17.5\leq i^{\prime} \leq 22.5$)
and an $i^{\prime}$-faint ($22.5 < i^{\prime} \leq 24.0$) subsample to
check the quality of the photometric redshifts. Also, we did our test
down to three different limits in $K$ magnitude ($K_{\rm{Vega}}\,\leq$
20.25, 20.5, 21.5), to evaluate the effects of the inclusion of
$K$-faint sources.  To ensure the use of reliable photometric
redshifts, we considered only those based on $\geq$ 5 photometric
bands (including the $K$ band).  Moreover, to avoid side-effects
related to the quality of spectroscopic redshifts, we limited the
comparison to galaxies with spectroscopic flags 3, 4, 23, and 24,
although we recognize that this choice could bias our subsamples
toward classes of objects whose spectral properties favour the
redshift measurement.  The results are summarized in
Table~\ref{zphot_zspec_comp}, where we report the values of
$\sigma_{\Delta z/(1+zs)}$ and $\eta$ for the selected subsamples of
galaxies and the total number of objects we used for our test.

\begin{figure*}
\includegraphics[width=9.5cm,clip=]{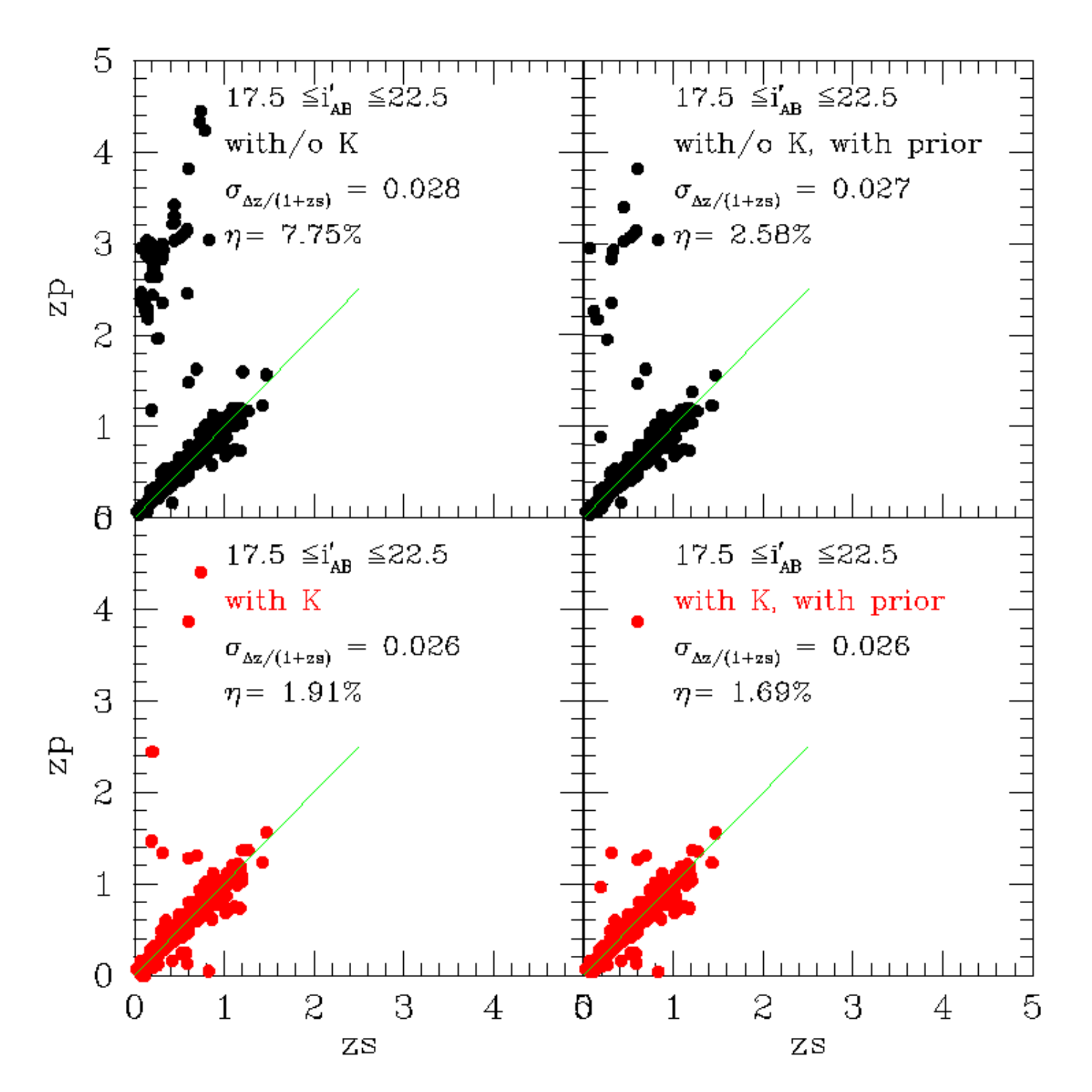}
\includegraphics[width=9.5cm,clip=]{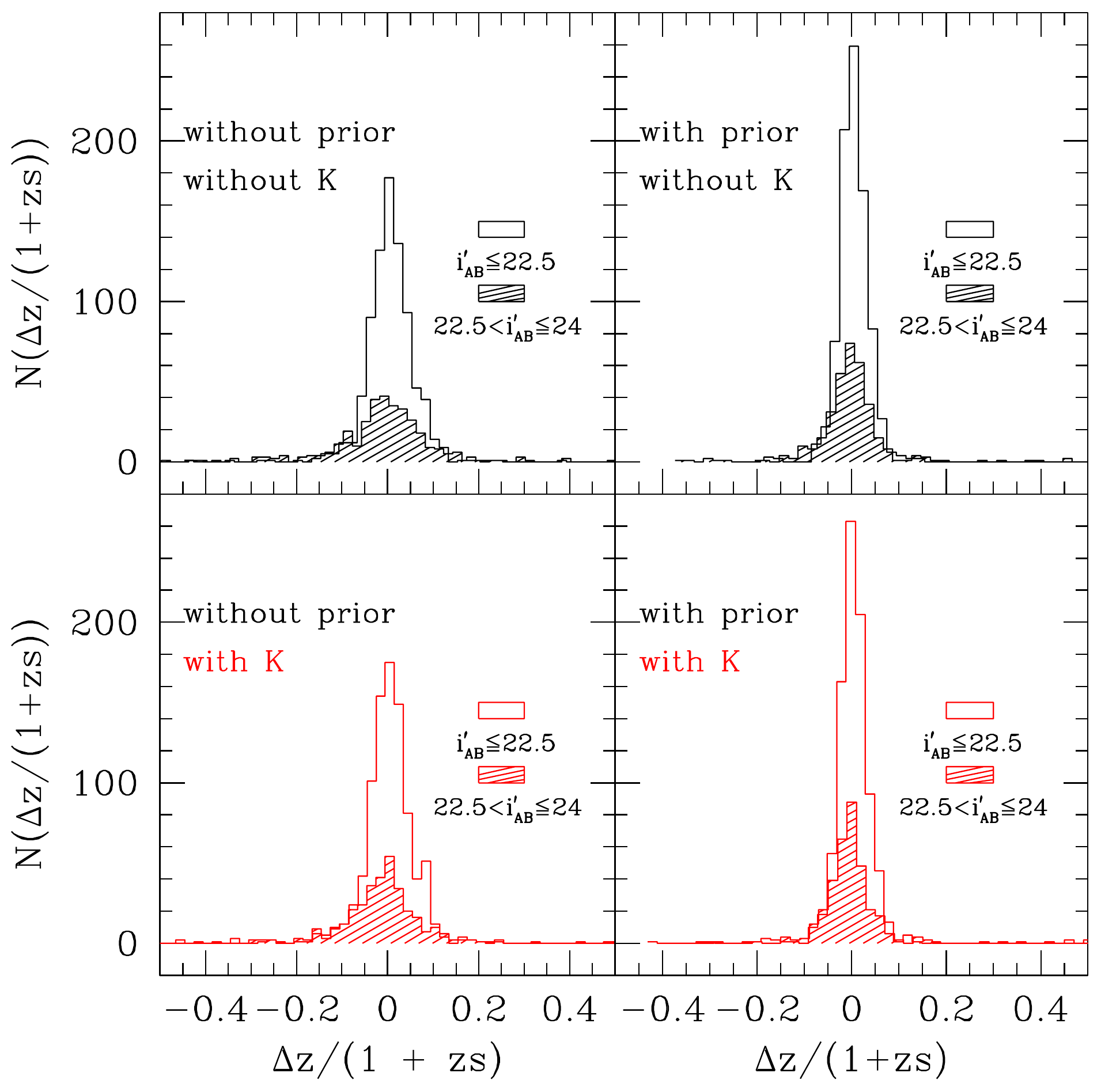}
\caption{Comparison between spectroscopic and photometric redshifts for the 
$i^{\prime}$-bright subsample (left-hand panels) and
distribution of $\Delta z/(1+zs)$ for the $i^{\prime}$-bright and $i^{\prime}$-faint
subsamples (right-hand panels, empty and shaded histograms, respectively). See text for details.
The $zp$ = $zs$ line is displayed in green.
The results for Le\_Phare fits with/without prior and with/without inclusion of $K$-band 
photometry are shown with the relevant values of $\sigma_{\Delta z/(1+zs)}$ and $\eta$. 
The results for the fits with the $K$-band are displayed in red. }
\label{zph_zsp}
\end{figure*}

Figure~\ref{zph_zsp} illustrates the effects of the inclusion of the
$K$-band and of the prior on the quality of photometric redshifts for
the $K$-selected spectroscopic sample down to $K_{\rm{Vega}}\, \leq\,
20.25$ (the results for the samples down to fainter $K$ magnitude
limits are alike).  It is evident that the inclusion of the $K$-band,
even without the use of any prior, is more effective than the use of
the prior alone in reducing the number of catastrophic redshifts,
especially for objects that are erroneously assigned a photometric
redshift in the range 2 $<\, zp \, \leq$ 3.5. The best result is
obtained with the use of both the $K$-band and the prior, which gives
$\sigma_{\Delta z/(1+zs)}$ = 0.026 and $\eta$ = 1.61\% for the
$i^{\prime}$-bright subsample, and $\sigma_{\Delta z/(1+zs)}$ = 0.028
and $\eta$ = 2.43\% for the entire sample, at least in the covered
redshift range.
  
Objects spectroscopically classified as active galactic nuclei are not
considered in the comparison shown here.  Their inclusion would
increment the catastrophic errors, but would leave unchanged the
outcome of the comparison, with the best quality of photometric
redshifts being achieved by including both the $K$-band and the prior
in the fitting process.

\begin{table*}
\begin{minipage}[t]{\textwidth}
\caption{Quality of photometric redshifts determined with/without $K$-band photometry and
with/without prior down to different $K$ magnitude limits and for two intervals of $i^{\prime}$ 
magnitudes. For the comparison we used galaxies
with good quality spectroscopic redshifts (flags 3, 4, 23, 24) and with photometric redshifts
derived from at least 5 photometric bands, including the $K$ band.}   
\label{zphot_zspec_comp}      
\centering
\renewcommand{\footnoterule}{}                         
\begin{tabular}{l c c c c c c}       
\hline\hline     
 & & & & & & \\
 & $K\leq 20.25$ & $K\leq 20.5$ & $K\leq 21.5$ & $K\leq 20.25$ & $K\leq 20.5$ & $K\leq 21.5$ \\
 Nobj & 890 & 917 & 946 & 347 & 453 & 817\\
 \hline
 & & & & & & \\ 
 & \multicolumn{3}{c}{$17.5\leq i^{\prime} \leq 22.5$} & \multicolumn{3}{c}{$22.5 < i^{\prime} \leq 24.0$}\\
 \hline
 & & & & & & \\ 
 & \multicolumn{6}{c}{without $K$, without prior}\\
 & & & & & & \\  
 \hline
 $\sigma_{\Delta z /(1+z)}$ & 0.0280 & 0.0280 & 0.0280 & 0.0384 & 0.0379& 0.0351 \\
 $\eta$ (\%) &   7.75  & 8.18 & 8.35 & 7.78 & 7.73 & 7.71 \\   
\hline
 & & & & & & \\ 
 & \multicolumn{6}{c}{ with $K$, without prior} \\
 & & & & & & \\ 
 \hline
 $\sigma_{\Delta z /(1+z)}$ &  0.0265 & 0.0265 & 0.0267 & 0.0349 & 0.0345& 0.0327 \\
 $\eta$ (\%) &   1.91 & 1.96 & 2.11 & 5.76 & 5.30 & 5.38\\
 \hline
 & & & & & & \\ 
 & \multicolumn{6}{c}{without $K$, with prior} \\ 
 & & & & & & \\ 
 \hline
 $\sigma_{\Delta z /(1+z)}$ & 0.0270& 0.0268 & 0.0268 & 0.0381 & 0.0367 & 0.0339 \\
 $\eta$ (\%) &    2.58 & 2.62 & 2.85 & 6.05 & 5.52 & 5.88 \\
 \hline
 & & & & & & \\ 
 & \multicolumn{6}{c}{with $K$, with prior} \\ 
 & & & & & & \\ 
 \hline
  $\sigma_{\Delta z /(1+z)}$ & 0.0264& 0.0264 & 0.0265 & 0.0350 & 0.0336 & 0.0322 \\
   $\eta$ (\%) &   1.69 & 1.75 & 1.90 & 4.32 & 3.97 & 4.65 \\ 
 \hline
\end{tabular}
\end{minipage}
\end{table*}

\begin{figure}
\includegraphics[width=\columnwidth,clip=]{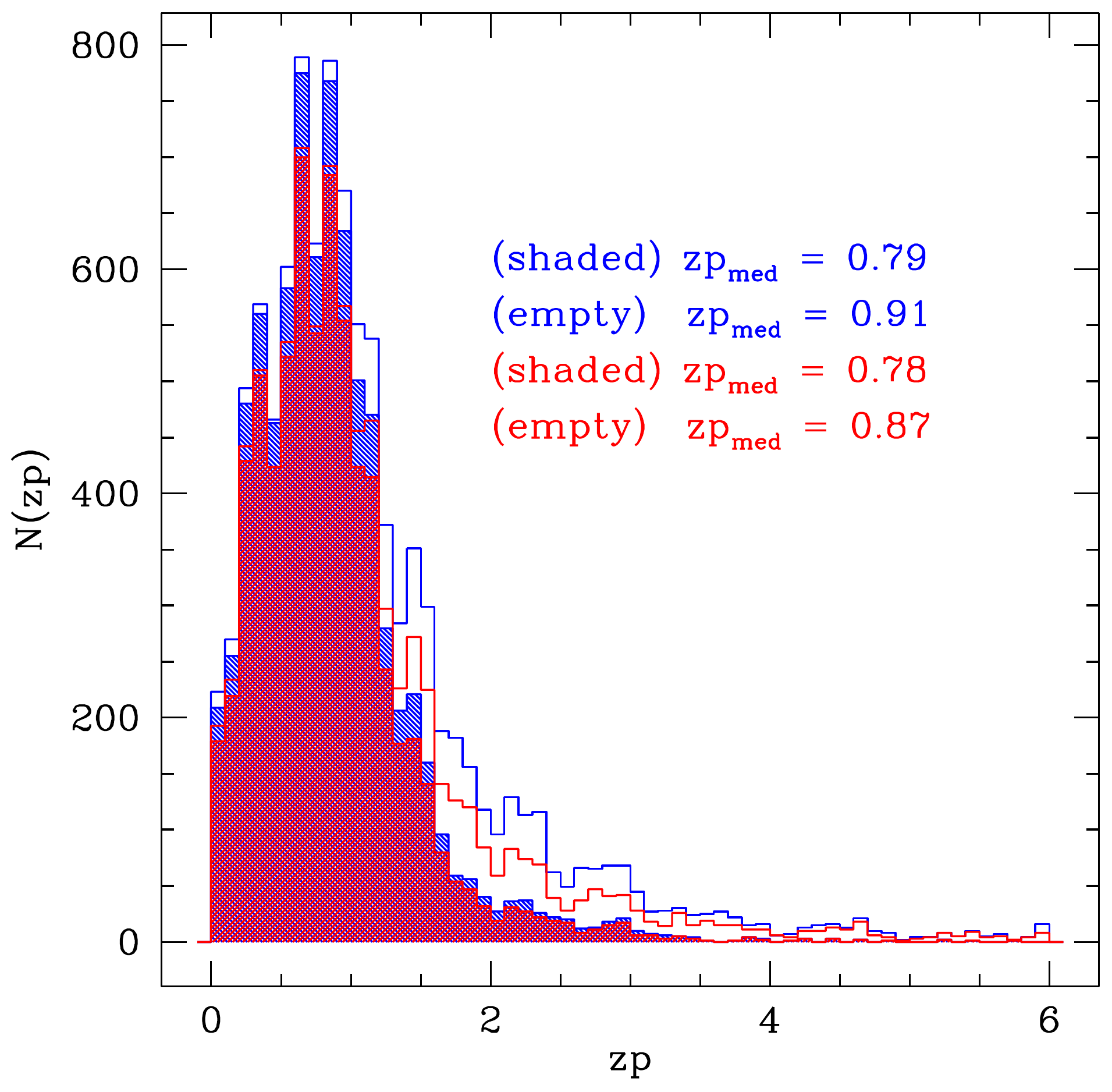}
\caption{Distribution of photometric redshifts for the $K$-selected
photometric samples down to $K_{\rm{Vega}}\, \leq 20.25$ (red) and 
$K_{\rm{Vega}}\, \leq 20.5$ (blue). Shaded histograms show the redshift distribution when 
an additional $I$-selection matching the VVDS selection
for the spectroscopic survey is applied. The median photometric redshifts of each 
distribution are reported.}
\label{Nphotoz}
\end{figure}

Because of the $I_{\rm{AB}}\, \leq\, 24$ magnitude limit of the sample
upon which the prior is based, we decided to adopt photometric
redshifts determined without the prior for sources fainter than this
limit.  In Fig.~\ref{Nphotoz} we show the photometric redshift
distribution of our sample, after the exclusion of candidate stars,
identified as described in Sect.~\ref{sg-sep}.  The redshift
distribution is shown for the two magnitude limits $K_{\rm{Vega}}\,
\leq 20.25$ and $K_{\rm{Vega}}\, \leq 20.5$ for a purely $K$-selected
sample and for a sample with an additional $I$-selection that
reproduces the magnitude limits of the VVDS spectroscopic survey. The
effect of the additional $I$-selection is the loss of objects
especially at intermediate to high redshift and, therefore, a lowering
of the median redshift of the sample.  The median redshift of the
purely $K$-selected photometric sample down to $K_{\rm{Vega}}\, \leq
20.25\, (20.5)$ is $zp_{med}$ = 0.87 (0.91).
 
The photometric redshifts determined in this work are used in a
companion paper that is dedicated to $K$-selected extremely red
objects \citep{st07} extracted from our $K$-band wide catalogue.

\section{Star-galaxy separation}\label{sg-sep}


We classified point-like objects with a combination of photometric and
morphological criteria, using parameters derived by SExtractor and
Le\_Phare.  We adopted a quite complex strategy since our aim is to
perform the star/galaxy separation even at faint magnitudes, where
most of the single criteria are known to fail.

The chosen parameters and adopted criteria are the following: 
\begin{itemize}
\item[i)] the CLASS\_STAR parameter given by SExtractor, providing
the ``stellarity-index'' for each object; selected point-like sources
are obeying CLASS\_STAR $\ge 0.95,0.90$ for objects brighter/fainter
then $i^{\prime}_{\rm AB} < 22.5$ respectively; 
\item[ii)] the FLUX\_RADIUS parameter, also
computed by SExstractor in the $K$-band and denoted as $r_{1/2}$, measuring the radius
that encloses $50$\% of the object total flux; we imposed $r_{1/2} <
3.4$ pixels as the criterion to be satisfied by stellar sources; 
\item[iii)] the MU\_MAX
parameter by SExtractor, representing the peak surface brightness
above the background; the locus of selection of stellar objects in the
$i'$ vs MU\_MAX plot has been derived empirically using the
spectroscopic sample; 
\item[iv)] the $BzK$ criterion, proposed by \citet{daddi04},
with stars characterized by colours $z-K < 0.3(B-z)-0.5$;
\item[v)] the $\chi^2$ and of the SED fitting carried out with Le\_Phare
during the photometric redshift estimate, by using template SEDs of both stars 
($\chi^2_{star}$)and galaxies ($\chi^2_{gal}$); 
the criterion to be satisfied by a stellar source is
$\chi^2_{gal} \, - \, \chi^2_{star} \, > $ 0.
\end{itemize}

\begin{figure*}
\centering
\includegraphics[width=16cm]{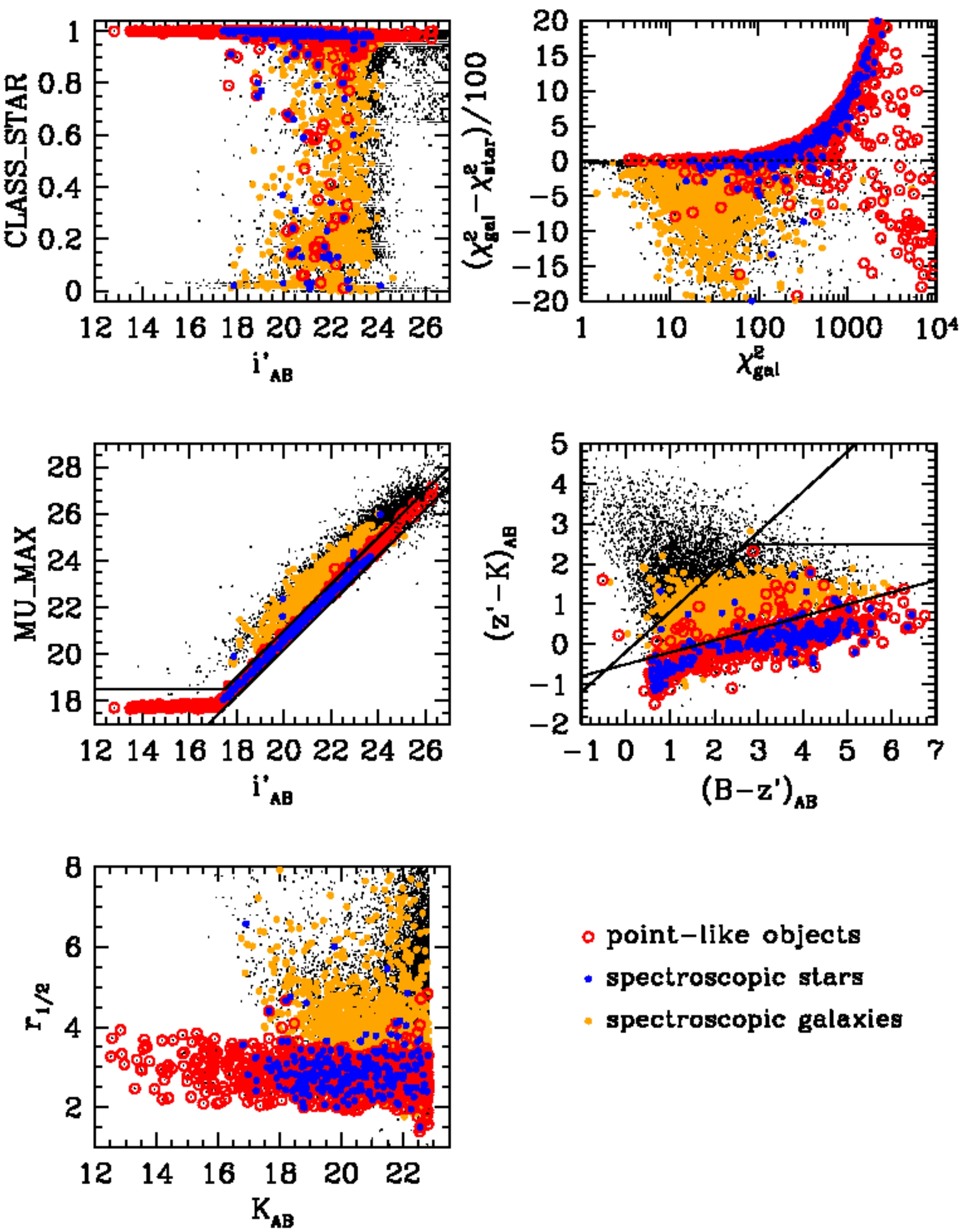} 
\caption{The five criteria for the star-galaxy separation (see text) applied to 
the spectroscopic sample. From top to bottom, and left to right, the SExtractor CLASS\_STAR,
 MU\_MAX, and FLUX\_RADIUS parameters, the Le\_Phare best-fit $\chi^2$ values, and the $BzK$
 diagram \citep{daddi04} are plotted. Each panel illustrates an individual photometric criterion 
and the corresponding candidate stellar sources (red circles), overplotted to
spectroscopically classified galaxies (yellow dots) and stars (blue dots). }
\label{sg-criteria}
\end{figure*}

The application of the above criteria to the spectroscopic sample is
shown in Fig.~\ref{sg-criteria}.  Objects fulfilling at least 4 of the
above mentioned criteria have been classified as point-like, whereas a
less restrictive constraint has been imposed for objects brighter than
$K_{\rm{Vega}}\, = \,17.16$, with only 3 over 5 criteria
fulfilled. Moreover, to be sure not to miss stellar saturated objects,
we included to the point-like objects selection all the sources with
$r_{1/2} < 4$ and $K_{\rm{Vega}}\, < \,14.16$ disregarding the other
criteria.

The selection yields 745 point-like objects at $K_{\rm{Vega}}\,\le
\,20.5$, with 235 of them belonging to the spectroscopic sample.
Within this subsample, 80\% of the spectroscopic stars are correctly
identified and only 3 spectroscopic galaxies and 4 active galactic
nuclei with very reliable redshift (flag= 3, 4, 13, 14) fall into the
star candidate subsample. The galaxy sample is found to be
contaminated by misclassified stars at the 2\% level.

This method for the star-galaxy separation represents an improvement
over both the method used by \citet{oi06} and the one adopted by
\citet{pozzetti07}, as we verified on the spectroscopic sample.

Our final classification of the objects takes into account the
spectroscopic information when this is flagged as highly reliable
(flag = 3, 4, 23, 24, 13, 14, 213, 214) by overriding the photometric
classification with the spectroscopic one.
 
\section{Galaxy and star number counts}\label{number-counts}

\begin{figure}
\centering
\includegraphics[width=9cm]{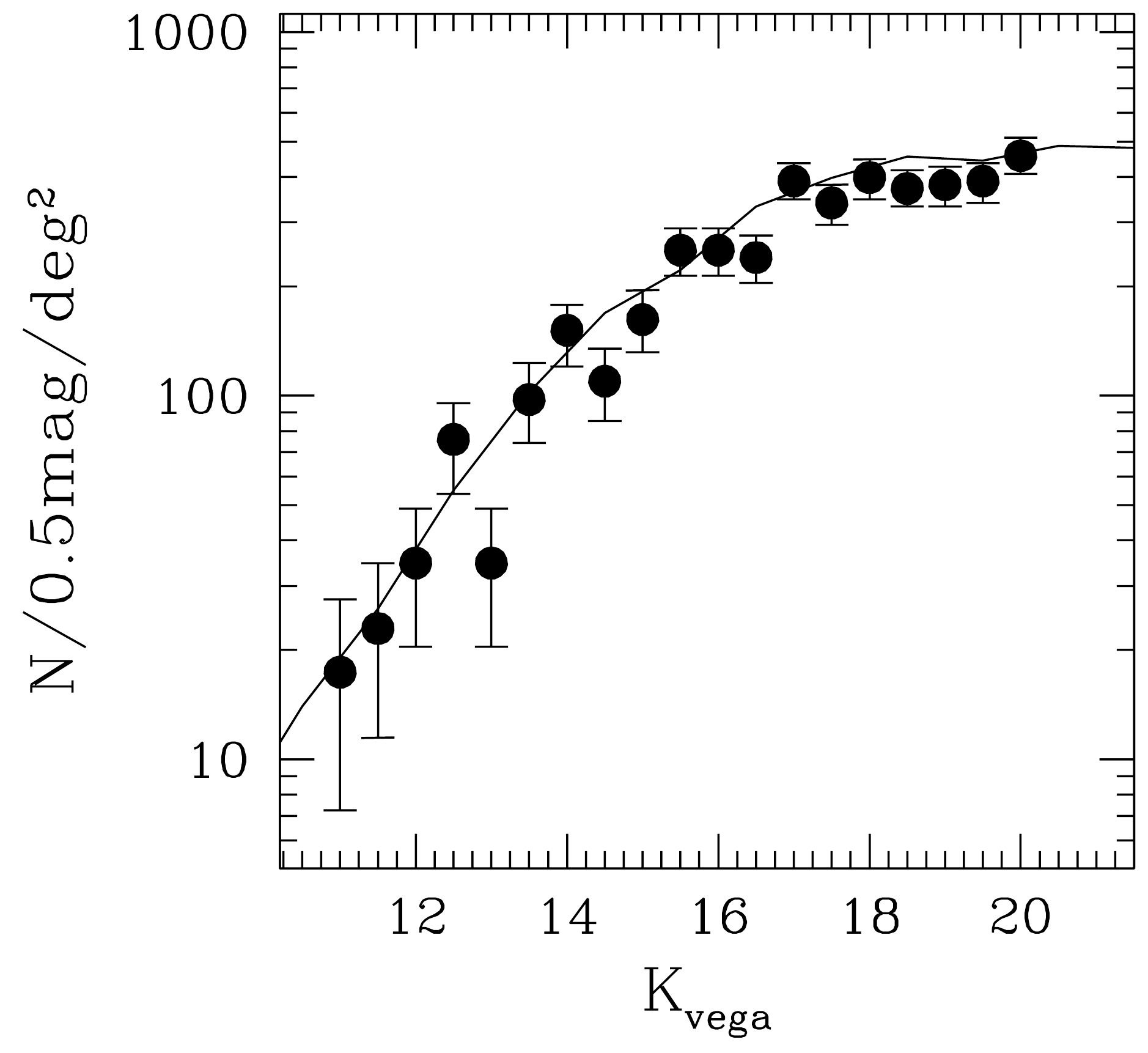} 
\caption{Differential number counts of stars in the F02 wide 
  field. The continuous lines are the prediction of the model of
  \citet{rob03}.}
\label{starcounts}
\end{figure}

Comparing number counts of galaxies and stars with published
compilations is a good check both of the star-galaxy separation
efficiency and of the reliability of our photometry, as well as the
sample reliability and completeness. The differential number counts of
stars (number 0.5~mag$^{-1}$ deg$^{-2}$) for the F02 wide field are
shown in Figure \ref{starcounts}.  To avoid underestimating
bright-star counts, for this exercise we used the catalogues before
excising the areas around bright stars. The continuous line is the
prediction of the model of \citet{rob03} computed at
the appropriate galactic latitude. The agreement between observed and
predicted star counts is very good (the error bars shown are
Poissonian error bars), confirming the reliability both of our
photometry and of our star-galaxy separation procedure.

\begin{figure}
\centering
\includegraphics[width=9cm]{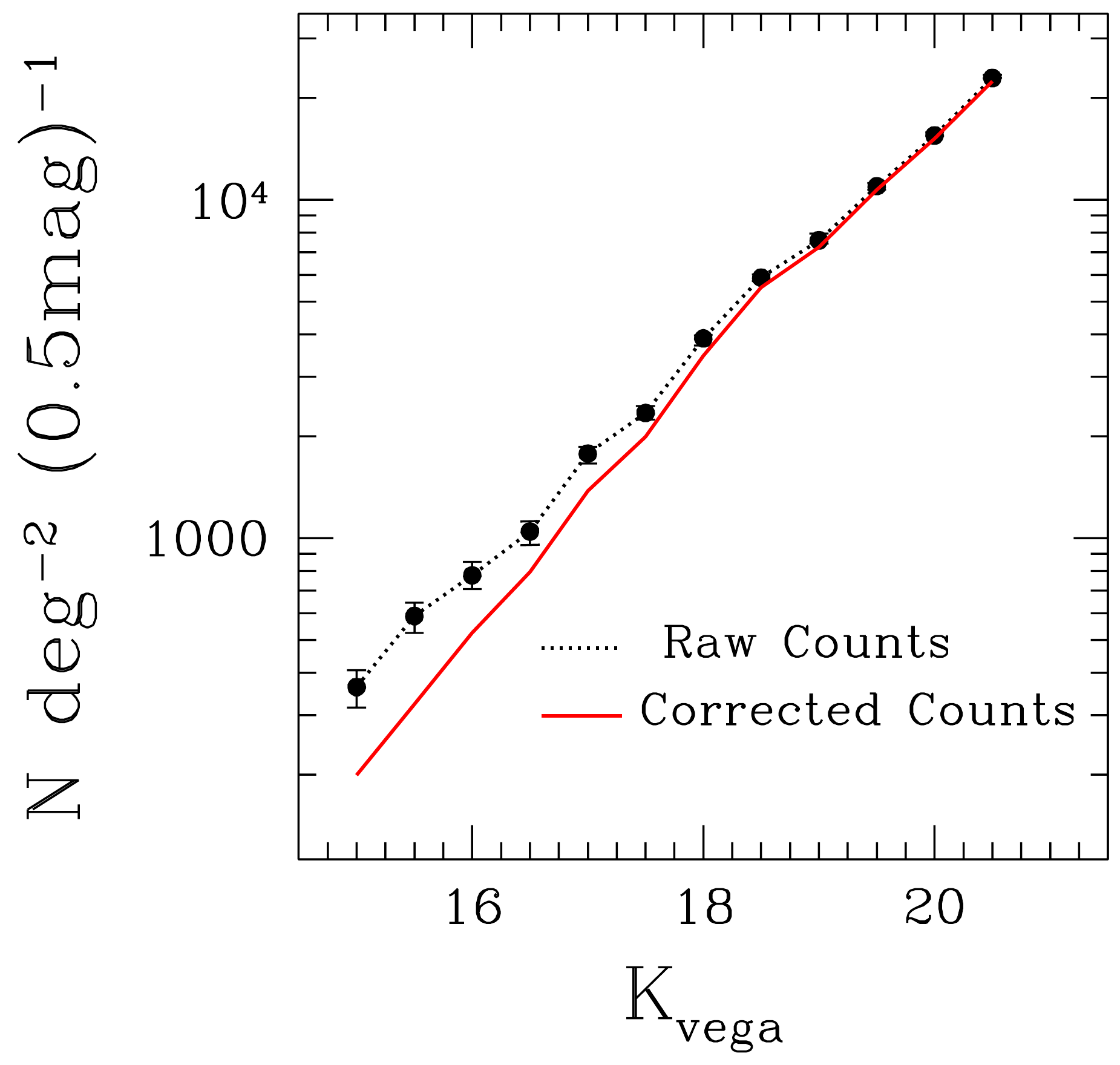} 
\caption{Differential $K-$band number counts in the F02 wide field.  
The dotted line shows the raw number densities and error bars are
Poissonian. The heavy line shows galaxy densities obtained after
correcting for star contamination. }
\label{galcounts}
\end{figure}

Figure \ref{galcounts} shows the differential number counts
(number~(0.5~mag)$^{-1}$~deg$^{-2}$) for the F02 wide field. The
dotted line is obtained by normalising the observed raw number counts
to the total field after excising areas around bright stars (616
arcmin$^2$).  The error bars shown are Poissonian error bars and no
correction for stellar contamination has been applied to the raw counts
shown. The contamination estimated from the prediction of the model of
\citet{rob03} is below 3\% for the fainter bins shown
in the plot. The heavy continuous line shows the total galaxy counts
obtained after correcting the counts for stellar contamination
estimated using our star-galaxy separation. We did not apply any
further completeness or contamination correction to our data because
down to the limiting magnitude ($K_{\rm{Vega}} \, =$ 20.5) plotted in Figure
\ref{galcounts} such corrections are negligible.

Table \ref{counts_K} lists our raw differential counts, the total raw
number densities (in units of number~(0.5~mag)$^{-1}$ deg$^{-2}$), and
the final, corrected for stellar contamination, galaxy densities
together with their $1\sigma$ error bars.

\begin{table}
        \caption{Column 2 shows the raw
differential number counts per half magnitude bin. Column 3 shows our
raw total number densities (number~0.5mag$^{-1}$ deg$^{-2}$). Column 4
shows our total, differential galaxy densities, corrected for star
contamination and Column 5 is their poissonian $1\sigma$ error bar.} 
\label{counts_K}
 \centering
  \begin{tabular}{cccccccc}
\hline
\hline
$K_{\rm{vega}}$& total    & {\it N$_{raw}$}               & {\it N$_{corr}$}              & $ \sigma $ \\
               & counts   & $0.5~$mag$^{-1}\rm{deg}^{-2}$ & $0.5~$mag$^{-1}\rm{deg}^{-2}$ & \\
\hline
15    &  62    &    363  &   199    &   34	\\ 	    
15.5  &  100   &    585  &   322    &   43	\\	  
16    &  133   &    778  &   520    &   55	\\	  
16.5  &  178   &   1041  &   801    &   68	\\	  
17    &  302   &   1766  &  1374    &   90	\\	  
17.5  &  402   &   2351  &  2006    &  108	\\	  
18    &  660   &   3859  &  3462    &  142	\\	  
18.5  &  1007  &   5889  &  5514    &  180	\\	  
19    &  1309  &   7655  &  7275    &  206	\\	  
19.5  &  1881  &  10999  & 10614    &  249	\\	  
20    &  2671  &  15619  & 15157    &  298      \\
            \hline
            \noalign{\smallskip}

          \end{tabular}
\end{table}

\begin{figure}
\centering
\includegraphics[width=9cm]{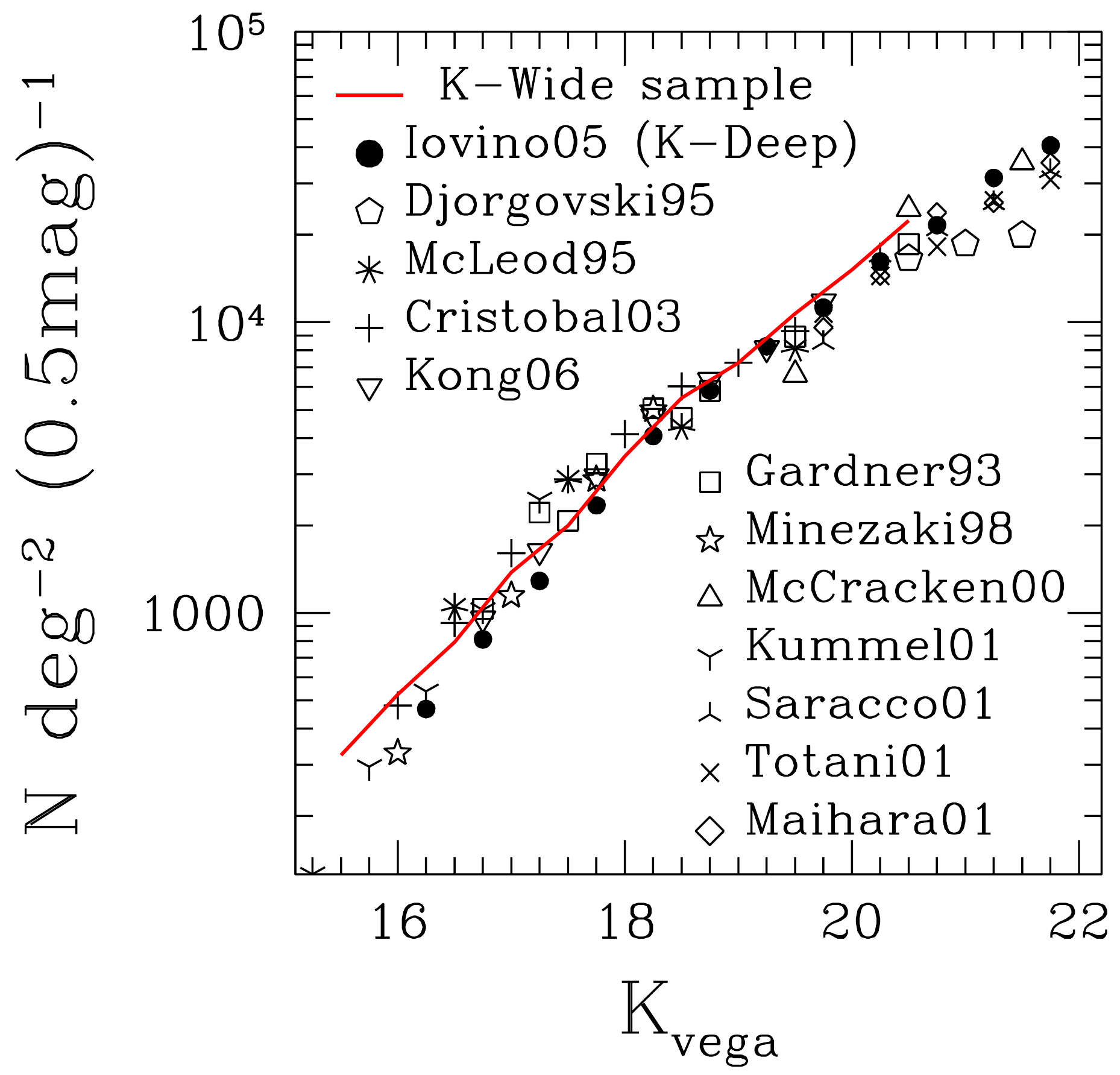}
\caption{The $K-$band galaxy number counts, in units of
number~0.5mag$^{-1}$ deg$^{-2}$, obtained in this paper are compared
to a compilation of those from the literature, including counts from
\citet{gard93}, \citet{dj95},
\citet{mcl95},
\citet{min98}, \citet{hjmc00}, 
\citet{kum01}, \citet{sar01},
\citet{tot01}, \citet{mai01},  
\citet{ch03}, \citet{iovino05}, 
and \citet{kong06}. }
\label{counts_lit}
\end{figure}

Figure \ref{counts_lit} show our total corrected galaxy
counts (solid line) compared with a selection of literature data. We
have followed the approach of \citet{ch03} and
select only reliable counts data from the literature, considering only
data with negligible incompleteness correction and with star-galaxy
separation applied. We have been conservative in the selection of the
magnitude intervals plotted in our counts, restricting ourselves to
bins with relatively large numbers of galaxies, negligible
incompleteness and small contamination corrections. The agreement with
literature data is very good. Our F02 wide galaxy counts confirm
the change of slope around $K_{\rm{Vega}}\, \sim \,18.0$ previously detected by
\citet{iovino05}, \citet{gard93} and 
\citet{ch03}. In the range $18\, < \, K_{\rm{Vega}} \, < \,20.5$ the 
slope of the galaxy counts is $\gamma_{K}\, \sim \, 0.30 \pm 0.09$, while
in the brighter magnitude range, $15.0 < K_{\rm{Vega}} < 18.0$, the slope is
steeper: $\gamma_{K}\, \sim \,0.42 \pm 0.02$.

\section{Clustering analysis for $K-$selected data}
In this Section we investigate the clustering properties of point like
and extended sources in the F02 wide $K-$band catalogues.

We use the projected two-point angular correlation function,
$\omega(\theta)$, which measures the excess of pairs separated by an
angle $\theta, \theta+\delta\theta$ with respect to a random
distribution. This statistic is useful for our purposes because it is
particularly sensitive to any residual variations of the magnitude
zero-point across our stacked images. We measure $\omega(\theta)$
using the standard \citet{ls93} estimator, i.e.,

\begin{equation}
\omega_{e} ( \theta) ={\mbox{DD} - 2\mbox{DR} + \mbox{RR}\over \mbox{RR}}
\label{1.ls}
\end{equation}

with the $\mbox{DD}$, $\mbox{DR}$ and $\mbox{RR}$ terms referring to
the number of data-data, data-random and random-random pairs between
$\theta$ and $\theta + \delta\theta$.  We use logarithmically spaced
bins, with $\Delta log(\theta) = 0.2$, and the angles are expressed in
degrees, unless stated otherwise.  $\mbox{DR}$ and $\mbox{RR}$ are
obtained by populating the two-dimensional coordinate space
corresponding to the different fields number of random points equal to
the number of data points, a process repeated 1000 times to obtain
stable mean values of these two quantities.

\subsection {Clustering of point-like sources} 

We first measure the angular correlation function $\omega(\theta)$ of
the stellar sources.  As stars are unclustered, we expect that, if our
magnitude zero-points and detection thresholds are uniform over our
field, then $\omega(\theta)$ should be zero at all angular scales.

\begin{figure}
\centering
\includegraphics[width=8.5cm]{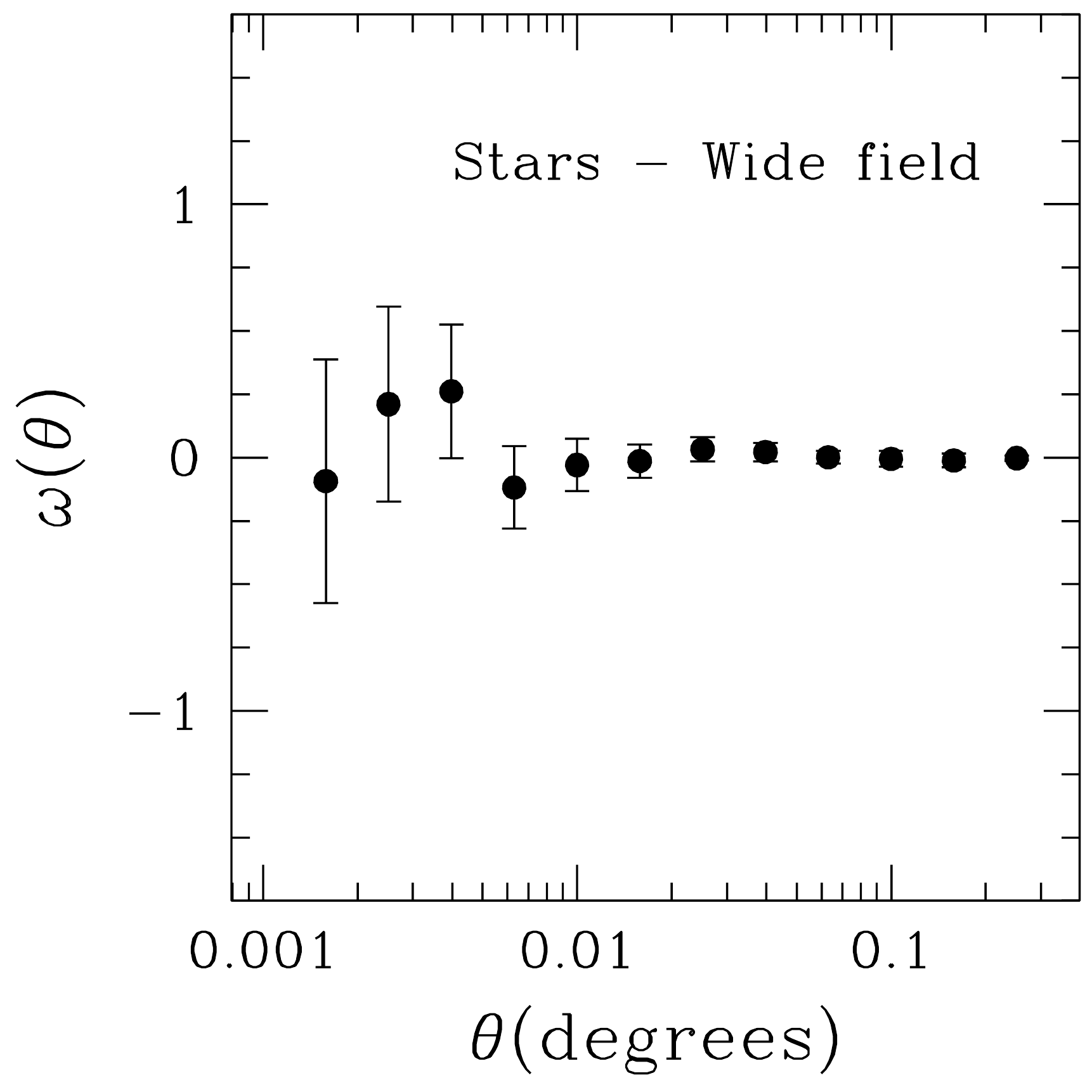} 
\caption{Plot of the correlation function $\omega(\theta)$ for the
total sample of stars in the $K-$band stacks as a function of the
logarithm of the angular pair separation in degrees. At all scales
displayed the measured correlation values are consistent with zero.
}
\label{ksi_stars}
\end{figure}

The results for F02 wide-field $K-$band data are displayed in
Figure \ref{ksi_stars}, where the correlation function is plotted for
the total sample of stars obtained from our wide field according to
the procedure described in section \ref{sg-sep}: 617 stars in
the $K-$band interval $15.00\, \leq\, K\, \leq\, 20.25$.  At all scales
displayed the measured correlation values are consistent with
zero. Error bars are obtained through bootstrap resampling of the star
sample (and are roughly twice poissonian error bars).

\subsection {Clustering of extended sources}
 
The procedure followed to measure $\omega(\theta)$ is similar to the
one described above for the star sample. In the case of galaxies a
positive amplitude of $\omega(\theta)$ is expected, and we have to take
into account the so called ``integral constraint'' bias.  If the real
$\omega(\theta)$ is assumed to be of the form
$A_{\omega}\theta^{\delta}$, our estimator (\ref{1.ls}) will be offset
negatively from the true $\omega(\theta)$, according to the formula:

\begin{equation}
\omega_e ( \theta) =A_{\omega}(\theta^{-\delta}-C)
\label{2.ls}
\end{equation}

This bias increases as the area of observation decreases, and it is
caused by the need to use the observed sample itself to estimate its
mean density, see {\it eg} \citet{peeb80}.  The
negative offset AC can be estimated by doubly integrating the assumed
true $\omega(\theta)$ over the field area $\Omega$:

\begin{equation}
A_{\omega}C = {1\over \Omega^2} \int\int \omega ( \theta) d\Omega_1 d\Omega_2
\label{3.ls}
\end{equation}

This integral can be solved numerically using randomly distributed
points for each field:

\begin{equation}
C = {\sum N_{rr}( \theta)~\theta^{-\delta} \over \sum N_{rr}( \theta)}
\label{4.ls}
\end{equation}

Assuming 1\arcsec as the pairs minimal scale at which two galaxies can be
distinguished as separated objects, and $\delta = 0.8$, we obtain the
following value: $C_{Wide} = 4.50$.

\begin{figure}
\centering
\includegraphics[width=8.5cm]{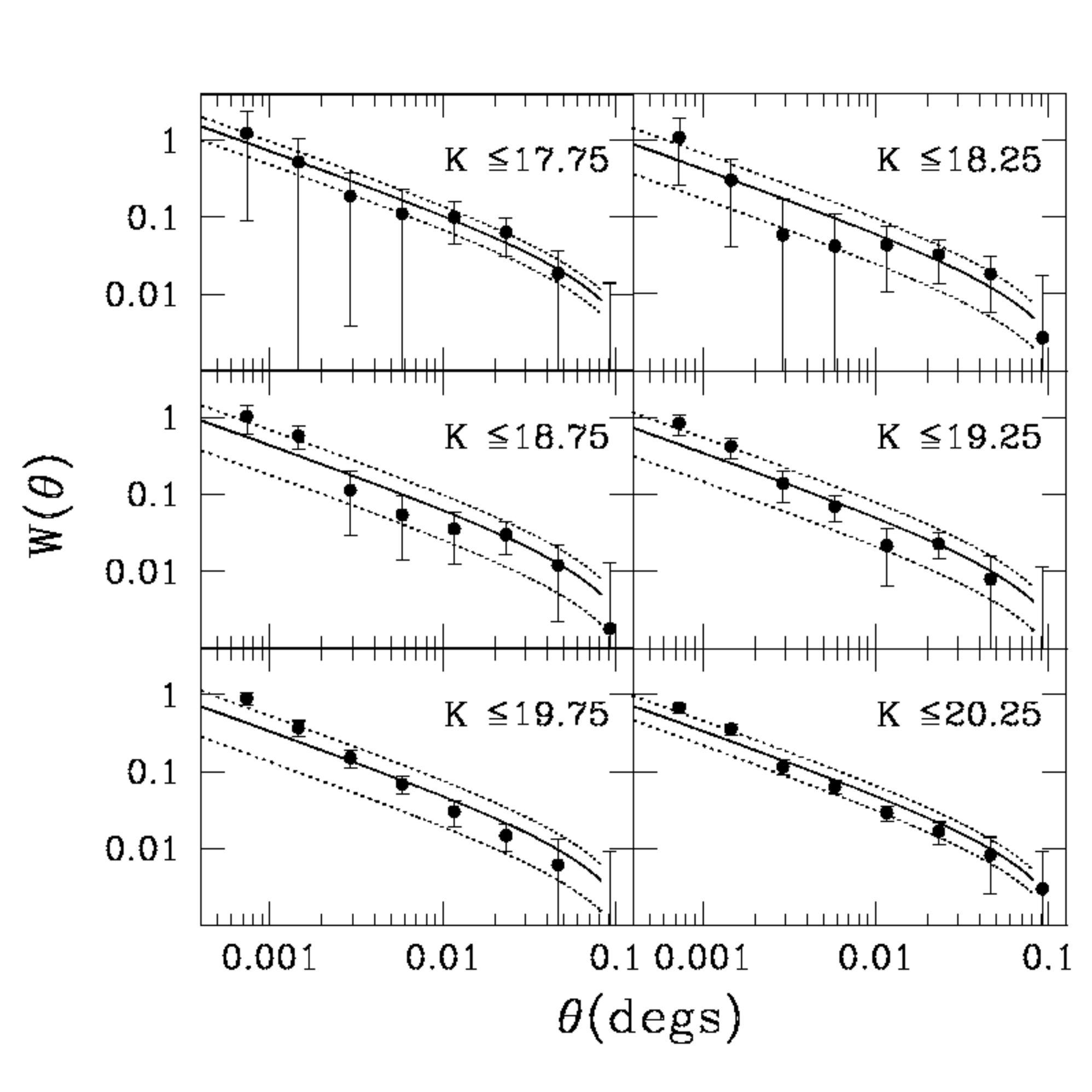} 
\caption{
Results obtained for the amplitude $A_{\omega}$ at $1\deg$ of our
F02 wide field on galaxy sub-samples of different $K-$band
limiting magnitudes. No correction for stellar contamination is
applied and the error bars on the amplitude are those obtained from
the fit. }
\label{ksi_wide}
\end{figure}

We estimated the amplitude $A_{\omega}$ for a series of $K$ limited
galaxy samples by least square fitting $A(\theta^{-0.8}-C)$ to the
observed $\omega(\theta)$, weighting each point using bootstrap error
bars.  Figure \ref{ksi_wide} shows the results obtained for 6
different $K-$band limiting magnitudes.  No correction for stellar
contamination is applied (only the objects classified as stars, using
the method described in section \ref{sg-sep}, were excluded from
the analysis) and the error bars on the amplitude are those from the
fit.

\begin{figure}
\centering
\includegraphics[width=8.5cm]{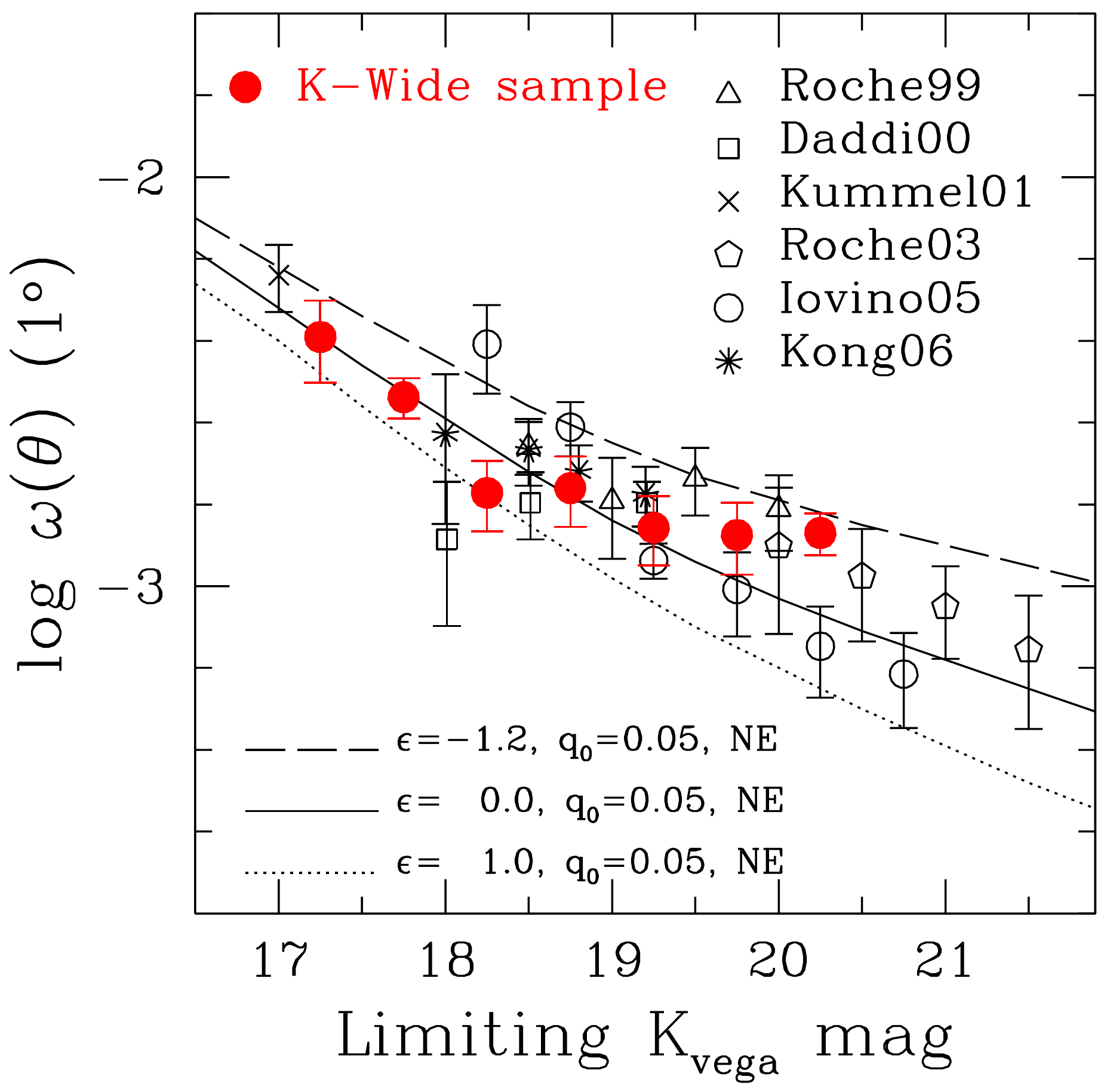} 
\caption{ We compare our total sample estimate of the amplitude,
$A_{\omega}$ at $1\deg$, of the angular correlation function with a
compilation of results from the literature in similar magnitude
ranges. We include mesurements from: \citet{roche99}, 
\citet{daddi00},
\citet{kum01},
\citet{roche03}, \citet{iovino05}, 
and \citet{kong06}.  The continuous, dotted and dashed
lines show, for reference, the models of PLE from
\citet{roche98}, with scaling from local galaxy
clustering. }
\label{ksi_literature}
\end{figure}

Figure \ref{ksi_literature} shows the comparison of our results with
literature data. The continuous, dotted, and dashed lines show the
models of PLE from \citet{roche98}, with scaling from
local galaxy clustering.  

Table \ref{amplitude} lists our amplitude measurements for the F02
sample, in units of $10^{-4}$ at $1\deg$, together with their
bootstrap error bars.  For each limiting magnitude, the total number
of objects N used in the analysis is also listed.
 
\begin{table}
        \caption{Observed $\omega(\theta)$ amplitudes A, in units of
$10^{-4}$ at $1\deg$, together with their $1\sigma$ error bars, for  our
F02 wide field. The number of objects used in the analysis down to the K magnitude faint limits
shown in the first column is also listed. 
}
        \label{amplitude}
 \centering
  \begin{tabular}{cccccccccc}
\hline
\hline 
Magnitude &\multicolumn {2}{c}{$0226-04$ (F02)} \\
          & N &A $\pm$ dA  \\
\hline 
K $<$ 17.25 &  561 & 40.68   $\pm$  9.24    \\
K $<$ 17.75 &  902 & 28.97   $\pm$  3.25    \\
K $<$ 18.25 & 1486 & 16.92   $\pm$  3.32    \\ 
K $<$ 18.75 & 2427 & 17.39   $\pm$  3.41    \\
K $<$ 19.25 & 3661 & 13.91   $\pm$  2.67    \\
K $<$ 19.75 & 5470 & 13.33   $\pm$  2.66    \\
K $<$ 20.25 & 8059 & 13.49   $\pm$  1.57    \\
            \hline
            \noalign{\smallskip}
          \end{tabular}
\end{table}

\section{Conclusions}

We have presented a new $K$-band catalogue that covers a contiguous
sky area of 623 arcmin$^2$ down to a magnitude limit $K_{Vega} \, \leq
\,21.5$ and provides us with a 90\% complete $K$-selected sample
down to $K_{Vega} \, \leq \,20.5$, although
with some possible level of color incompleteness potentially affecting
extremely red sources fainter than $K_{Vega} \, = \,20.5$, whose
inclusion is disfavoured by the procedure to build the catalogue at
these faint magnitudes (see discussion in Sect.~\ref{compl}).  This is
one of the biggest $K$-selected samples available to date to this
magnitude limit and is complemented by
$UBVRIu^{\ast}g^{\prime}r^{\prime}i^{\prime}z^{\prime}$ photometry --
available through the VVDS and CFHTLS surveys -- as well as by VVDS
spectroscopy with a sampling rate of $\sim$ 27\%, down to the 90\%
completeness limit.

Good quality photometric redshifts have been obtained for the whole
sample by following the method outlined by \citet{oi06} and including
the $K$-band photometry in the fits of the galaxy SEDs.  By using our
spectroscopic subsample we explored the effects of the inclusion of
$K$-band photometry in the determination of photometric redshifts.  We
verified that the use of the $K$ band leads to similar advantages as
the training on the spectroscopic redshift distribution function,
namely a considerable reduction of the fraction of catastrophic
errors.  When no a priori spectroscopic information is adopted to
train the fitting procedure, the additional use of the $K$ band
improves significantly the determination of photometric redshifts, as
expected.  The results from this survey show that we expect a
significant improvement in the accuracy of photometric redshifts
obtained by future wide-field surveys using near-infrared data
(e.g. the Visible and Infrared Survey Telescope for Astronomy,
VISTA\footnote{http://www.vista.ac.uk}).

Also, taking advantage of the $K$-band photometric parameters, we have
implemented a quite complex set of criteria for the star-galaxy
separation that improved on methods previously adopted by our team
\citep{oi06,pozzetti07}.  A very good agreement of star and galaxy
number counts with those present in the literature has proven the
effectiveness of our star-galaxy separation method as well as the good
quality of our photometry and the reliability and completeness of our
sample.

The good quality of our photometry as well as the reliability and
completeness of the sample are confirmed by the comparison of the
number counts of stars and galaxies with published compilations.  The
$K$-band galaxy number counts from this work are in excellent
agreement with those obtained from the $K$-deep sample
\citep{iovino05} and with a selection of compilations from the
literature.  The projected two-point angular correlation function does
not show any peculiarity as a function of magnitude and angular scale
and is broadly in agreement with results from the literature.\\

The galaxy mass function, the luminosity function, and the properties
of extremely red objects based on our $K$-selected catalogue are
presented in companion papers
\citep{pozzetti07,bolzonella07,st07,bondi07b}.  A cross-match of this
$K$-band catalogue with radio sources from the VVDS-VLA deep field has
been carried out by \citet{bondi07a} and has revealed a higher
incidence of faint $K$-band counterparts ($K_{\rm{Vega}} \, > \, 20$)
among candidate ultra-steep-spectrum radio sources with respect to the
rest of the radio sources in the sample, although based on small
number statistics.

\begin{acknowledgements}
This research has been developed within the framework of the VVDS
consortium.
We are grateful to E. Pompei who provided us with 
additional NTT $K$-band observations, needed to 
complete a pointing, in February 2006.
We thank the anonymous referee for his/her constructive
comments that helped to make the paper clearer.
This work has been partially supported by the
CNRS-INSU and its Programme National de Cosmologie (France),
and by Italian Ministry (MIUR) grants
COFIN2000 (MM02037133) and COFIN2003 (no.2003020150).
The VLT-VIMOS observations have been carried out on guaranteed
time (GTO) allocated by the European Southern Observatory (ESO)
to the VIRMOS consortium, under a contractual agreement between the
Centre National de la Recherche Scientifique of France, heading
a consortium of French and Italian institutes, and ESO,
to design, manufacture and test the VIMOS instrument.    
\end{acknowledgements}

\end{document}